\documentclass[twocolumn,amsmath,amssymb,prb]{revtex4}

\usepackage{graphicx}
\usepackage{dcolumn}
\usepackage{bm}
\usepackage{color}
\usepackage{ulem}
\usepackage{amsfonts,amsthm}
\usepackage{amssymb,slashbox}

\begin{document}

\def\vec#1{\boldsymbol #1}
\newcommand{\bR}{\mbox{\boldmath $R$}}
\newcommand{\tcy}[1]{\textcolor{black}{#1}}
\newcommand{\tn}[1]{\textcolor{black}{#1}}
\newcommand{\tm}[1]{\textcolor{black}{#1}}
\newcommand{\tr}[1]{\textcolor{black}{#1}}
\newcommand{\trr}[1]{\textcolor{black}{#1}}
\newcommand{\tmm}[1]{\textcolor{black}{#1}}
\newcommand{\tmmm}[1]{\textcolor{black}{#1}}
\newcommand{\tb}[1]{\textcolor{black}{#1}}
\newcommand{\tbb}[1]{\textcolor{black}{#1}}
\newcommand{\tg}[1]{\textcolor{black}{#1}}
\newcommand{\tgg}[1]{\textcolor{black}{#1}}
\newcommand{\tgf}[1]{\textcolor{black}{#1}}
\newcommand{\Ha}{\mathcal{H}}
\newcommand{\mh}{\mathsf{h}}
\newcommand{\mA}{\mathsf{A}}
\newcommand{\mB}{\mathsf{B}}
\newcommand{\mC}{\mathsf{C}}
\newcommand{\mS}{\mathsf{S}}
\newcommand{\mU}{\mathsf{U}}
\newcommand{\mX}{\mathsf{X}}
\newcommand{\sP}{\mathcal{P}}
\newcommand{\sL}{\mathcal{L}}
\newcommand{\sO}{\mathcal{O}}
\definecolor{green}{rgb}{0,0.5,0.1}
\newcommand{\la}{\langle}
\newcommand{\ra}{\rangle}
\newcommand{\ga}{\alpha}
\newcommand{\gb}{\beta}
\newcommand{\gc}{\gamma}
\newcommand{\gs}{\sigma}
\newcommand{\vk}{{\bm{k}}}
\newcommand{\vq}{{\bm{q}}}
\newcommand{\vR}{{\bm{R}}}
\newcommand{\vQ}{{\bm{Q}}}
\newcommand{\vga}{{\bm{\alpha}}}
\newcommand{\vgc}{{\bm{\gamma}}}
\newcommand{\Ns}{\mbox{N_{\text{s}}}}
\newcommand{\xx}{\mbox{$x^2-y^2$}}
\newcommand{\zz}{\mbox{$z^2$}}
\renewcommand{\figurename}{{\bf Figure}}
\renewcommand{\thefootnote}{\arabic{footnote}}
\setcounter{footnote}{0}


\title{ 
\tr{ Superconductivity and its mechanism 
in an $ab$ $initio$ model for electron-doped LaFeAsO}
}

\author{Takahiro Misawa and Masatoshi Imada}
\affiliation{ 
Department of Applied Physics, University of Tokyo, 7-3-1 Hongo, Bunkyo-ku, Tokyo, 113-8656, Japan
}


\begin{abstract}
\noindent  
{ Two families of high temperature superconductors whose critical temperatures are higher than 50K are known. One is the copper oxides and the other is the iron-based superconductors.  
Comparisons of mechanisms between these two in terms of common ground as well as distinctions will greatly help in searching for higher $T_c$ superconductors. 
However, studies on mechanisms 
for the iron family 
based on first principles calculations are few.
Here we first show that the superconductivity emerges in the 
state-of-the-art numerical calculations for an {\it ab initio} multi-orbital model of 
an \tcy{electron-doped} iron-based superconductor {LaFeAsO}, 
in accordance with experimental observations.
Then the mechanism of the superconductivity is 
\tgg{identified as \tcy{enhanced} uniform density fluctuations} by one-to-one correspondence 
with the instability toward  inhomogeneity driven by  
first-order antiferromagnetic  and nematic 
transitions. Despite many differences, certain common features with the 
copper oxides are also figured out in terms of 
the underlying \tr{orbital selective Mottness} found in the iron family. }
\end{abstract}

\maketitle

Discovery of the iron-based superconductors 
in 2008 opened a new way to reach 
high temperature superconductors~\cite{kamihara2008}.
It was found that there exist many similarities with
the other \tr{high-$T_{\rm c}$} superconductors, namely the copper oxides; 
superconductivity occurs in 
iron (copper) layers
upon carrier doping to the antiferromagnetic (AF) phase.
(In a typical iron-based superconductor LaFeAs(O,F), the AF order 
is destroyed at the electron doping concentration $\delta\sim 0.05$, 
above which the superconductivity takes over.)

However, significant dissimilarities also exist;
one of the significant is the active orbital degrees of freedom\tr{;} 
five Fe $d$ orbitals are involved near the Fermi surface in the iron-based superconductors, while
in the copper oxides, only one Cu $d$ orbital mainly constitutes the Fermi surface.
Thus, one central issue of the iron-based superconductors is the role of 
orbital degrees of freedom.

Because of the nearby 
AF phase, many theoretical and experimental studies 
proposed that spin fluctuations play an essential role  in stabilizing the 
superconductivity in common with proposals in copper oxides
\cite{KurokiPRL,MazinPRL,ChubukovPRB2008,Scalapino2009NJP,Mukuda2014PRB89_064511,PlattPRB2011}.
{In particular, Platt $et$ $al.$~\cite{PlattPRB2011} discussed roles of magnetic fluctuations
based on the first principles study combined with the functional renormalization group.}
Meanwhile, roles of 
orbital or nematic fluctuations were also suggested
~\cite{KontaniOnariPRL,Fernandes2014}. 
Although the normal
state properties including magnetism were numerically analysed successfully  
in the iron-based superconductors~\cite{ZPYin,misawa2012},
the superconductivity has not been studied by quantitative {\it ab initio} calculations.
Despite a large amount of works~\cite{IshidaJPSJReview,StewartRMP2011,ScalapinoRMP,PlattHankeThomale2013AdvPhys},
the primary cause that controls the \tr{high-$T_{\rm c}$} superconductivity is still unresolved. 
Identification of the mechanism requires treating the spin and orbital degrees 
of freedom on an equal footing together with precise temporal and spatial quantum fluctuations.

{In the present study, we numerically study the {\it ab initio}
low-energy effective models for the iron-based superconductors, particularly for
electron doped LaFeAsO by unprecedentedly large-scale computations, 
from which we identify superconducting mechanism in LaFeAsO.} 

{Our first interest is the significance of electron 
correlations~\cite{JYang2009PRL,Qazilbash2009NP,Degiorgi2011NJP,Terashima2010JPSJ,Nakai2008JPSJ}.  In particular, the issue is whether the present {\it ab initio} study reproduces 
{the proximity of the orbital-selective 
Mott insulator~\cite{Anisimov_2002,KogaPRL2004,Pruschke_2005, 
Arita_2005,Medici_2005,Ferrero_2005,Costi_2007,Jakobi_2009,Medici_2011,Greger_2013} 
}, 
as was pointed out in the literature \tgg{for the iron-based superconductors} both from 
theoretical and experimental 
analyses~\cite{IshidaLiebsch,Tamai,Aichhorn,misawa2012,MediciCapone,ZPYin,Yi_PRL,Lanata_PRB,Hardy_PRL,Li_PRB,Yu_2013,Wang_2014}. 
Simpler perspective by a two-orbital model also exists~\cite{Kou_EPL,Hackl_NJ,Yin_PRL,Zhang_PRB}.
It was established that a specific orbital $d_{X^2-Y^2}$, pinned close to half filling upon electron doping near the mother compound, LaFeAsO, 
shows a proximity to an orbital-selective Mott 
insulator\cite{misawa2012}. This nearly Mott-localized $d_{X^2-Y^2}$ orbital develops the AF order, where 
it couples to other orbitals by the Hund's rule coupling and forms the high-spin moment.
The origin of the orbital-selective Mottness on the {\it ab initio} grounds was interpreted from the higher 
density of states of the $d_{X^2-Y^2}$ orbital at the Fermi level in the original bare band structure\cite{misawa2012}. 
{The higher density of states makes the orbital more sensitive to the interaction effect.} We particularly focus on this orbital-selective behavior as an
underlying electronic structure that induces the superconductivity.}

{Then the next important issue we examine is whether the
superconductivity and its symmetry are correctly reproduced in the {\it ab initio} models for electron doped LaFeAsO. The final and central issue is the mechanism of the superconductivity.}

{Here, we demonstrate first that the orbital selective Mott insulating behavior 
indeed emerges in our {\it ab initio} calculations for undoped LaFeAsO.
Then we show that the electron doping  
eventually causes the depinning from the nearly Mott insulating $d_{X^2-Y^2}$ orbital into metals triggering 
first-order AF transitions between high and low spin states.
Filling-controlled first-order transitions generically drive inhomogeneity and phase separation. 
We next find that the superconductivity emerges in essential agreement with the experiments\cite{kamihara2008,Mukuda2014PRB89_064511}.
The pairing has a full gap and satisfies the symmetry 
with the opposite sign between the Fermi 
{pockets} at the $\Gamma$ and $M$ points in the Brillouin zone 
(so called $s\pm$ symmetry)
\cite{KurokiPRL,MazinPRL}.
Finally we show} by controlling the model parameters that the region of the strong 
density fluctuation near the phase separation remarkably has one-to-one correspondence with the 
identified superconducting regions. This smoking-gun observations support 
that the superconductivity in the iron-based superconductors is 
induced by the uniform charge {(or in other words electron density)} fluctuations discussed in the literature.
The density fluctuation from
the stripe-type AF order necessarily involves attraction of opposite spins 
at the next-neighbor bonds, which also generates the $s\pm$ singlet pairing. 
Thereby generated superconductivity is stabilized by increased coherence of the $d_{X^2-Y^2}$ carriers.
Common and distinct features offer insights into the copper oxides as well. 

\noindent
{\bf Results.} \\
{\bf Model derivation and framework.}
In showing the smoking gun of the superconductivity, 
we are based on an $ab$ $initio$ two-dimensional electronic model for
LaFeAsO derived by using the downfolding procedure,
which is detailed in the literature~\cite{miyake2010}.
The {\it ab initio} Hamiltonian $\mathcal{H}$ has the kinetic part $\mathcal{H}_{0}$ and the 
interaction part $\mathcal{H}_{\rm int}$ as 
\begin{align}
\mathcal{H} &= \mathcal{H}_{0}+\mathcal{H}_{\rm int},  \label{Eq:Ham1}\\
\mathcal{H}_{0}&=\sum_{\sigma} \sum_{i,j} \sum_{\nu,\mu}  
  t_{i,j,\nu,\mu} 
                   c_{i,\nu,\sigma}^{\dagger} 
                   c_{j,\mu,\sigma},   \label{Eq:Ham1_b}\\
\mathcal{H}_{\rm int} &= \mathcal{H}_{\text{\rm on-site}}+\mathcal{H}_{\text{\rm off-site}}.
\label{Eq:Ham2}
\end{align}
Here, $c_{i,\nu,\sigma}^{\dagger}$ ($c_{i,\nu,\sigma}$)
creates (annihilates) an electron with
spin $\sigma$ on the $\nu$th Wannier orbital 
at the $i$th site.
$t_{i,j,\nu,\mu}$ contains single-particle levels and transfer integrals.
Details of $\mathcal{H}_{\rm int}$ and $\mathcal{H}_{0}$ including transfer integrals and  interaction parameters such as
on-site intra-orbital/inter-orbital Coulomb interactions 
and exchange interactions 
are found in {Methods}, Supplementary {Tables I and II}, and
in the literature~\cite{miyake2010,nakamura2010,misawa2012}.

The present model contains five Fe $3d$ orbitals \tr{such as} 
$d_{XY}$, $d_{YZ}$, $d_{Z^{2}}$, $d_{ZX}$ and $d_{X^2-Y^2}$.
(We note that the ($X,Y$) axis are rotated by 45$^{\circ}$ from $(x,y)$ 
directed to the
Fe-Fe direction (e.g., the direction of the 
$d_{XY}$ orbital is parallel to the nearest Fe-Fe direction.).

To analyze the ground state of the $ab$ $initio$ model, we employ the 
many-variable variational Monte Carlo 
(mVMC) method~\cite{TaharaVMC_Full} (see also Methods for details), 
which appropriately takes into account the strong correlation effects
after considering both the quantum and spatial fluctuations. 
All the calculated extensive physical quantities are shown as those per site and the unit of the energy 
is eV while the length unit is the nearest Fe-Fe distance. 
Detailed definitions of magnetic as well as superconducting 
quantities studied in the present work are given in Methods. 

\begin{figure}[htb!]
  \begin{center}
    \includegraphics[width=8cm]{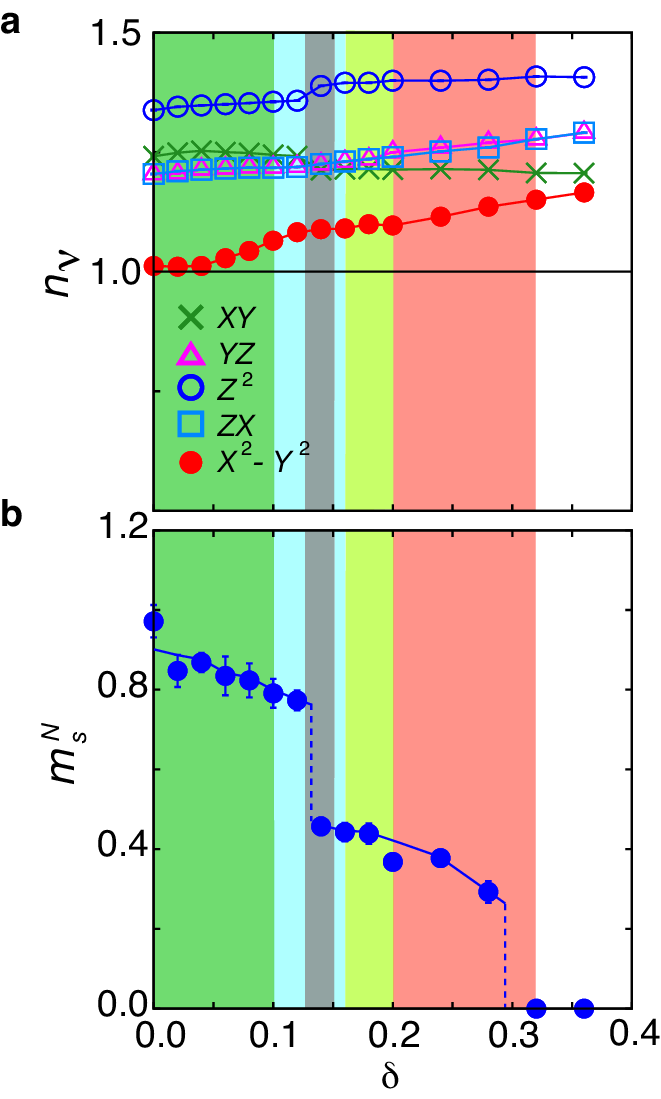}
  \end{center}
\caption{
~{\bf 
\tr{Orbital occupancies and AF order parameter} of {\it ab initio} electronic model.}
For definitions of colored areas (phase diagram) and physical     
quantities, see also Fig.~\ref{fig:Fig2} and Methods.  
{\bf a,} Orbital \tn{resolved} filling as functions of doping concentration $\delta$.
{\bf b,} $\delta$ dependence of stripe-type AF order parameter 
for the normal state $m_{s}^N\equiv m_{s}({\bm q}=(0,\pi))$ (filled blue circles),
which indicates the AF order with large $m_{s}^N$ (LAF) (dark green) 
and
small $m_{s}^N$ (SAF) (light green) 
phases.
In all the figures hereafter,
the error bars indicate the statistical
errors of the Monte Carlo sampling. Curves are guides for the eye.
}
\label{fig:Fig1}
\end{figure}%

\begin{figure}[htb!]
  \begin{center}
    \includegraphics[width=8cm]{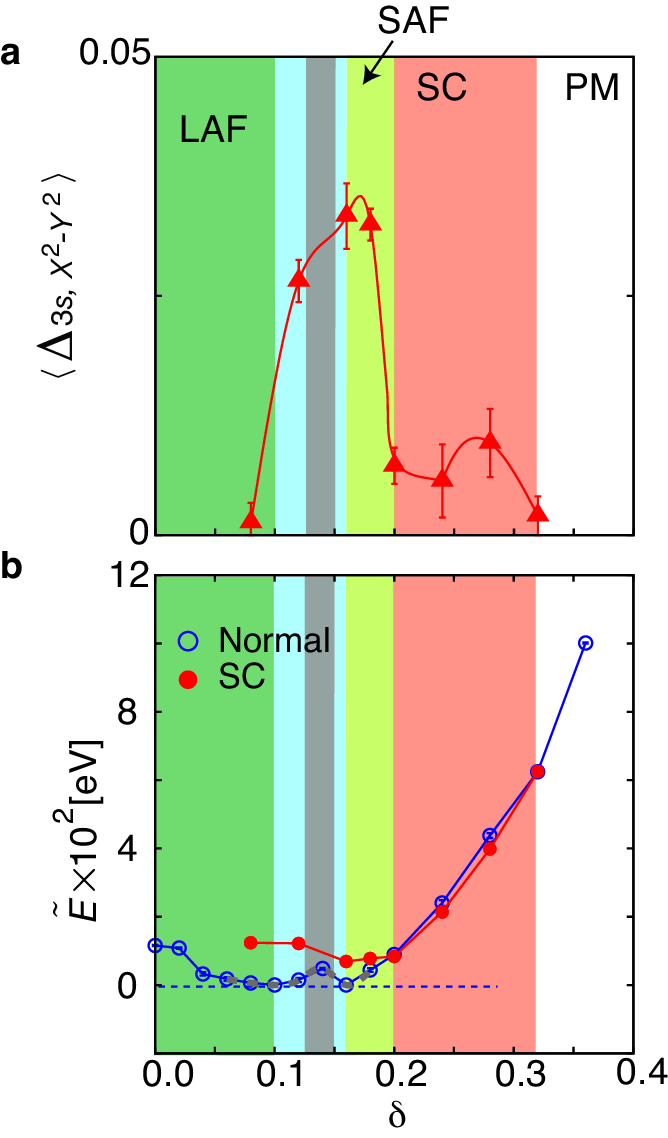}
  \end{center}
\caption{
~{\bf \tr{Doping dependence of superconducting order parameter and total energies}}
{\bf a,}
{$s$-wave  (gapped $s_{\pm}$)
superconducting order parameter of the $d_{X^{2}-Y^{2}}$ 
orbital $\langle \Delta_{3s,X^{2}-Y^{2}} \rangle$ in the superconducting phase.} 
It is conspicuous that $\langle \Delta_{3s,X^{2}-Y^{2}} \rangle$ 
has peaks around the first-order jumps of $m_s^N$.
\tr{SC indicates the superconducting phase.}
{\bf b,}
{ $\delta$ dependence of total energy
$\tilde{E}=E/N_{\rm s}-f(\delta)$ of normal (blue circles)
and superconducting (red circles) states.} 
The ground state is superconducting for
$0.2<\delta<0.32$ (red area).
We subtracted a common linear term $f(\delta)=a+b\delta$ ($a$ and $b$ are constants)
from the bare total energy per site $E/N_s$ for clarity.
The spinodal region (see Methods) 
estimated as
$0.12\leq\delta\leq0.15$ (gray region) 
 is obtained by
fitting around $\delta=0.13$ up to the fourth polynomial (the bold gray broken curve).
The phase separation region ($0.1<\delta <0.16$ shown as light-blue area) is 
determined by the Maxwell's construction (\tn{blue} dashed line).
The notations are the same as Fig.~\ref{fig:Fig1}.
}
\label{fig:Fig2}
\end{figure}%

\noindent
{\bf Fully $ab$ $initio$ electronic model.}
Now, we show results obtained by solving the fully $ab$ $initio$ electronic model. 
Figure~\ref{fig:Fig1}a shows the doping concentration ($\delta$) dependence 
of the occupation of each orbital $n_{\nu}$ for the normal state. Here, $\delta=0$ corresponds 
to the undoped mother compound LaFeAsO, where six $3d$ electrons 
occupy an iron site on average. It clearly shows that the $d_{X^2-Y^2}$ orbital (red filled circle) is
pinned close to half filling $n_{X^2-Y^2}\sim 1$ upon electron doping 
from $\delta=0$ up to $\sim$0.05 and the doping is small until the 
first-order transition at $\delta\sim 0.13$, indicating that the $d_{X^2-Y^2}$ orbital 
stays nearly at the orbital selective Mott insulator.   
The orbital selectivity in iron-based superconductors was  
discussed in several theoretical works~\cite{misawa2012,IshidaLiebsch,Aichhorn,ZPYin,MediciCapone,
Yi_PRL,Lanata_PRB,Hardy_PRL,Li_PRB,Kou_EPL,Hackl_NJ,Yin_PRL,Zhang_PRB}.
\tm{
We also find that the  $d_{X^2-Y^2}$ has the smallest double occupancies (see Supplementary \tr{Fig.~1a}),
which indicates that the $d_{X^2-Y^2}$ is located near the
orbital selective Mott insulator.}
The mechanism of the orbital differentiation \tm{in the $ab$ $initio$ model} was discussed 
in Ref. \onlinecite{misawa2012} as we mentioned above. 
The proximity of the orbital-selective Mott insulator is indeed 
confirmed {in the present {\it ab initio} model at $\delta=0$ as $\chi_{cX^2-Y^2}=0.008(10), \chi_{cXY}=0.32(4), \chi_{cZ^2}=0.27(4), \chi_{cYZ}=0.26(3), \chi_{cZX}=0.25(3)$, 
where $\chi_{c\nu}\equiv dn_{\nu}/d\mu$ is the orbital-dependent charge compressibility. 
The compressibility $\chi_{c\nu}$ is 
prominently small for $\nu=X^2-Y^2$ as is expected 
from the proximity of the orbital selective Mott insulator.}
Accordingly, near $\delta=0$, it develops the stripe-type AF order as we see in Fig.~\ref{fig:Fig1}b, 
where the largest contribution to the ordered moment $m^N_s$ comes from the $d_{X^2-Y^2}$ 
orbital {(see Supplementary \tr{Fig.~1b} )}
as is naturally expected as the Mott insulating nature 
of this orbital. Other orbitals are dragged by the $d_{X^2-Y^2}$ orbital to 
the ordered state realizing the high-spin state thanks to the Hund's rule 
coupling $J_H$ (We confirmed that the AF order disappears even at $\delta=0$ if 
orbital-dependent $J_{H\nu\mu}$ is uniformly reduced to $85\%$ of the {\it ab initio} value).  

The electron doping destroys the AF order similarly to the copper oxides. 
However, the suppression is initially slow because of the high-spin state, 
where the \tr{orbital blocking}~\cite{ZPYin} does not seem to allow a 
gradual evolution of the kinetic energy gain.
{However, with the electron doping, this suppression of the kinetic energy gain is 
released by the transition to the lower-spin state, when the kinetic energy gain 
exceeds the energy gain by the Hund's rule coupling. 
It easily occurs as a first-order transition as in other high-spin low-spin transitions. 
This mechanism can be tested by directly calculating the off-site Green functions, 
which reflect the kinetic energy. The kinetic energy gain indeed suddenly 
increases at the first-order transition. As an example, we show the 
doping dependence of the {nearest-neighbor} orbital-diagonal 
Green functions in Supplementary \tr{Fig.~2}. We find that $X^2-Y^2$ component of the 
Green's function drastically increases around the first-order phase transition, which is consistent with above mechanism.}

{Indeed the} high- to low-spin transition  
drives two successive first-order phase transitions; 
the first one occurs around $\delta\sim0.13$ from large-$m_{s}$ AF (LAF)  phase
to small-$m_{s}$ AF (SAF) phase.
The second one occurs around $\delta\sim 0.3$ between SAF and paramagnetic (PM) phases. 
As the stable phase, the 
second first-order transition is preempted by the emergence of the 
superconducting phase as we discuss later.

We also confirmed that the nematic order that breaks the four-fold rotational 
symmetry to the two-fold one coexists in the AF phase and shows a 
strong first-order transition simultaneously with the AF transition.
\tcy{After the first-order phase transition, 
we find that the nematic order seems to remain finite even in the PM phase.}
This may correspond to the electronic nematic phase observed in
BaFe$_{2}$(As$_{1-x}$P$_{x}$)$_2$ and 
Ba(Fe$_{1-x}$Co$_{x}$)$_{2}$As$_{2}$~\cite{Chu2010,Kasahara2012}, 
though it is not clear in LaFeAs(O,F) for the moment.

In addition to the normal state, 
we now examine superconductivity: 
We started from various 
\tr{Bardeen-Cooper-Schrieffer (BCS)} type superconducting wavefunctions 
that have different symmetries as the initial condition,
and then relax and optimize all the variational parameters to lower the energy.
After the optimization, we find that only the superconducting 
state with the gapped $s_{\pm}$-wave symmetry, which we call $3s$ hereafter, survives.
For detailed
definitions  of the 
superconducting states with $3s$ symmetries and others, see Methods. 
In Fig.~\ref{fig:Fig2}a, we show, by (red) triangles, the pairing order parameter of 
the gapped $s_{\pm}$-wave symmetry contributed from the $d_{X^{2}-Y^{2}}$ orbital 
($\langle \Delta_{3s,X^{2}-Y^{2}} \rangle$) determined from the long-ranged 
(leveled-off) part of the pairing correlations, if it saturates to a nonzero value.
Because the $d_{X^{2}-Y^{2}}$ orbital gives the dominant 
contribution to $\langle \Delta_{3s}\rangle$ and contributions from other orbitals are small,
we only show $\langle \Delta_{3s,X^{2}-Y^{2}} \rangle$.
{The possibility of orbital-selective superconductivity 
has been discussed in the literature~\cite{Yu_2014}.} 

This result indicates that the $d_{X^{2}-Y^{2}}$ orbital
governs the superconductivity as well as magnetism. 
In the sense that the single orbital 
($d_{X^{2}-Y^{2}}$) plays a dominant role in
stabilizing the superconductivity, 
it is similar to that of the copper oxides.
The proximity to 
the half filling and its departure process of electrons in the $d_{X^2-Y^2}$ orbital is a key to 
stabilize the superconducting phase.

However, Fig.~\ref{fig:Fig2}a also shows a crucial difference from the copper oxides. 
The superconducting order parameter shows dome structures near the strong first-order transition between LAF and SAF (as well as SAF and PM),
where neither the AF nor the orbital (nematic) fluctuations show enhancement. At the first-order transition, the antiferromagnetic and nematic order parameters show a simple jump between the two rather doping independent values, which clearly show that their fluctuations are small. Therefore, the present {\it ab initio} result is difficult to reconcile with the spin/orbital fluctuation mechanism as the glue of the pairing. We will further inspect this correspondence between the first-order transition and the dome peak later.

In Fig.~\ref{fig:Fig2}b, we show $\delta$ 
dependence of the total energies $E_{\rm n}$ and $E_{\rm s}$ for 
the normal (open blue circles) and superconducting (filled red circles) phases, respectively. 
(Both are at least at local minima of the free energy for $0.08\leq \delta\leq 0.32$).
For $0.2<\delta<0.32$ (red region), the 
superconducting state becomes the true ground state because $E_{\rm s}<E_{\rm n}$. 

\begin{figure}[htb!]
  \begin{center}
    \includegraphics[width=8cm]{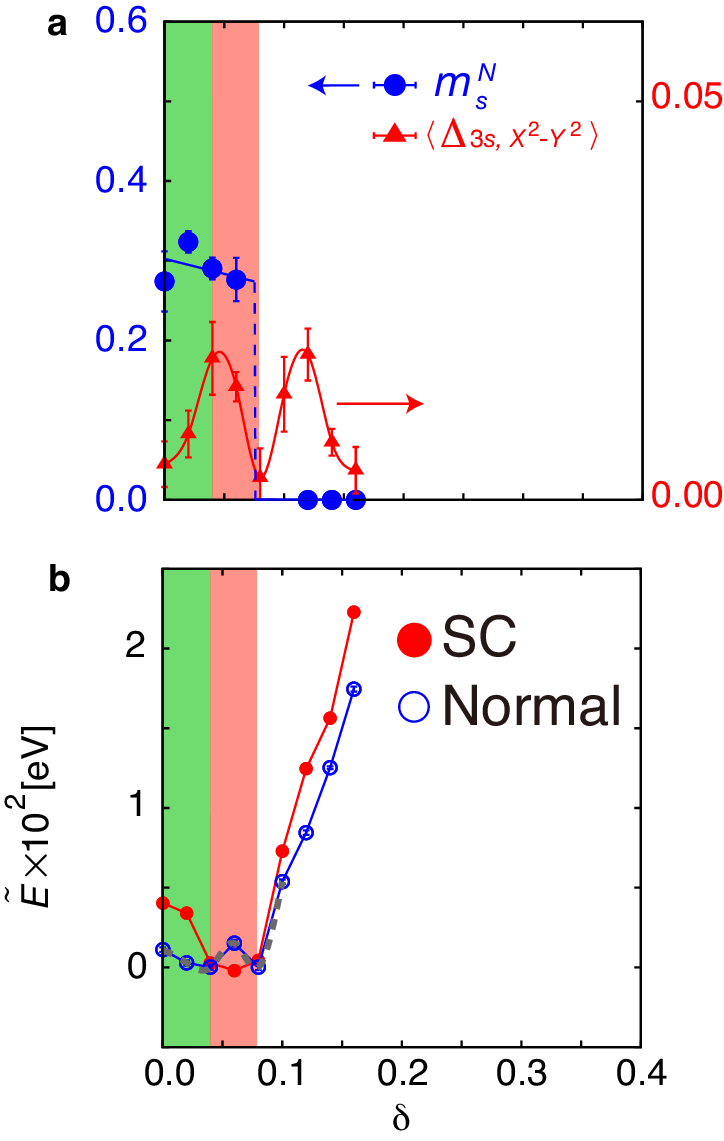}
  \end{center}
\caption{
{\bf 
\tr{Doping dependence of AF order parameter, superconducting order parameter,
and total energies} at $\lambda=0.95$.} {\bf a,} Doping concentration dependence 
of stripe AF order $m_{s}^N$ in the normal state \tn{(blue circles)} and the
superconducting order parameter in the superconducting state \tn{(red triangles)}.
{\bf b,} Total energies {per site} for normal and superconducting phases
as functions of $\delta$. The notations are the same as Figs.~\ref{fig:Fig1} and \ref{fig:Fig2}.
}
\label{fig:Fig3}
\end{figure}%

From $\delta$ 
dependence of the total energy,
the Maxwell's construction,
given by the broken thin blue straight line, 
determines the region of the phase separation $0.1<\delta<0.16$, where the 
thermodynamically stable and uniform phase is prohibited. 
Around the first-order transition at $\delta=0.13$, {the phase separation necessarily occurs when the filling is controlled because the phase coexistence occurs at the same chemical potential. The phase separation depicted by 
light blue and gray areas is an inevitable consequence of the first-order transition.} 
Instabilities toward the phase separation are indeed 
experimentally observed in LaFeAsO$_{1-x}$F$_{x}$~\cite{LangPRL2010}, 
and Ba$_{1-x}$K$_{x}$Fe$_{2}$As$_{2}$~\cite{ParkPRL2009,InosovPRB2009}  but in smaller regions. 
{In addition, K$_{x}$Fe$_{2-y}$Se$_{2}$~\cite{Li_2012}, 
and Rb$_{x}$Fe$_{2-y}$Se$_{2}$~\cite{Texier_2012} 
suggests phase separation into iron-vacancy-ordered and 
iron-vacancy-free regions, possibly driven by the \tgf{underlying} electronic 
phase separation into antiferromagnetic and superconducting regions.}
{In terms of the present result, it is intriguing to examine in more detail the uniformity of the other 
iron-based superconductors that do not show clear evidences for the phase separation so far.}
{A small spinodal region (grey region) is seen, where the 
second derivative of energy with respect to density is 
negative meaning the thermodynamically unstable region (see Methods).} 
In the region 
between the phase separation (light blue) and the superconducting (red) regions, 
we find a stable SAF (light green) region that has a small moment $(m^N_s \sim 0.4)$.

The emergence of the superconducting dome upon electron doping 
after the destruction of the stripe-type AF order qualitatively 
and essentially reproduces the experimental 
phase diagram~\cite{kamihara2008,Mukuda2014PRB89_064511}. 
Therefore, it is now desired to clarify the origin of this superconductivity. 

Before discussing the superconducting mechanism, however, 
it should be noted that the obtained phase diagram looks 
quantitatively different from the experimental phase diagram
of LaFeAs(O,F) in which AF and phase separation are limited to smaller doping regions.
We discuss the origin of these discrepancies below.\\

\begin{figure}[tb!]
  \begin{center}
    \includegraphics[width=8cm]{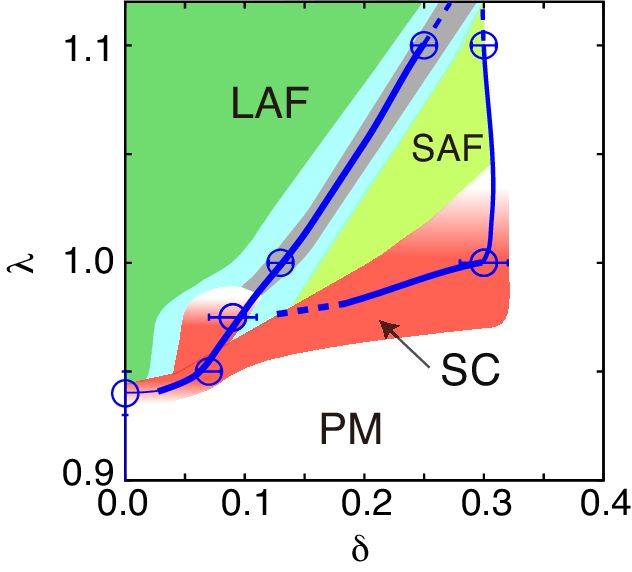}
  \end{center}
\caption{
{\bf Phase diagram in $\lambda-\delta$ plane.} Here, $\lambda$ is the
parameter to scale the interaction energy and $\delta$ is the doping concentration.
At $\delta=0$, AF order parameter continuously vanishes
at \tn{$\lambda\sim 0.94$}.
Red region represents
the superconducting phase, while
the LAF and SAF phases are drawn as dark and light green areas, respectively.
Gray area represents the spinodal region sandwiched by the
light-blue phase separation region. Blue curves represent the magnetic transition    
lines in the normal phase, which are preempted when it is located
in the spinodal (gray) or superconducting (red) regions.
The transition is mostly of first order {(bold curve)} except for the
region very close to $\delta=0$, where it looks continuous.   
}
\label{fig:Fig4}
\end{figure}%

\noindent
{\bf Effects of controlling electron correlations.} 
To understand correlation effects on the phase diagram {in more general},
we monitor the interaction by introducing 
the uniform scaling parameter $\lambda$ only for the interaction part as 
\begin{align}
\mathcal{H}=\mathcal{H}_{0}+\lambda\mathcal{H}_{\rm int}.
\end{align}
Namely, $\lambda=1$ represents the original $ab$ $initio$ 
electronic model. 

For \tn{$\lambda=0.95$}, as a function of $\delta$, the magnetic 
ordered moment $m_s$ and superconducting {order parameter} $\langle \Delta_{3s,X^2-Y^2}\rangle$ 
are plotted  in Fig.~\ref{fig:Fig3}a
and  the total energy is shown in Fig.~\ref{fig:Fig3}b. 
The dome of $\langle \Delta_{3s,X^2-Y^2}\rangle$ again appears near the first-order transition.
However, it turns out that
a small change \tn{of $\lambda$} induces a drastic
change, where 
the AF region largely shrinks to $0<\delta<0.04$ and the phase separation region in the normal state is
replaced with the stable superconducting phase in the region $0.04<\delta<0.08$.

In Fig.~\ref{fig:Fig4}, we draw the  global phase 
diagram in the $\lambda$-$\delta$ plane obtained by extensive calculations.
It is noteworthy that the experimental phase diagram of LaFeAs(O,F) is
consistent with that at a parameter $\lambda$ between 0.95 and 1.0 in terms of the regions of the magnetic stripe order with the moment
$m_s^N\sim 0.7$ and the superconducting dome. The strong first-order transition between the stripe antiferromagnetic and superconducting phases at $\delta\sim 0.05$ is also consistent with the experimental phase diagram.  In LaFeAs(O,F), the phase diagram shows an orthorhombic-tetragonal structural transition nearly at the magnetic-superconducting transition.  Since the electron-lattice coupling is not considered in the present {\it ab initio} model, the structural transition cannot be reproduced. Nevertheless, the present result supports that the orthorhombic-tetragonal transition is driven by the transition of the nematic order accompanied by the magnetic transition, which supports that the first-order transition is driven by the electronic mechanism. 

The consistency with the experimental result at $\lambda\sim 0.97$  implies a slight 
 ($<5\%\sim 0.1$ eV) overestimate in the {\it ab initio} values of the interaction, 
which could arise from the possible error in the downfolding procedure, 
and we conclude the essential agreement between the calculated result with the experiment.
 Furthermore, beyond the present  electronic {\it ab initio} scheme, such a small reduction of the effective interaction may arise from the electron-phonon interaction, 
where the frequency-dependent effective attraction was estimated as 0.4 eV but only within the range of the Debye frequency $\sim 0.02$eV~\cite{NomuraPRL}.  

The AF phase disappears and the superconductivity emerges in the mother compound LaFePO~\cite{Kamihara_LaFePO}. 
This is again consistent with the present phase diagram since the {\it ab initio} model of 
LaFePO corresponds to $\lambda<1$ and $\delta=0$~\cite{misawa2012}.

By increasing $\lambda$ 
beyond 1.0, the
AF phase becomes 
 \tn{quickly}  wider 
up to $\delta=0.3$. 
This sensitivity to the interaction
may account for recent experimental results of 
LaFeAsO$_{1-x}$H$_{x}$, where the AF phase reappears
in the overdoped region $\delta\gtrsim 0.4$~\cite{Iimura2012,Fujiwara2013PRL}.
Actually, it is reported that hydrogen substitution 
increases the anion height around $\delta\sim0.4$~\cite{Yamaura}.
The increase enhances the effective interactions because 
the screening from the anion $p$ orbitals becomes poorer~\cite{miyake2010}.
The reappearance of the AF phase in LaFeAsO$_{1-x}$H$_{x}$
accounted in this way is an interesting future subject of the first principles study.
\\

\noindent
{\bf Control of off-site interactions.}
To get further insight into the superconducting mechanism, 
let us study the {\it ab initio} model but here by switching off the off-site interactions $V_{nn}$ and $V_{nnn}$. 

\begin{figure}[t!]
  \begin{center}
    \includegraphics[width=6.5cm]{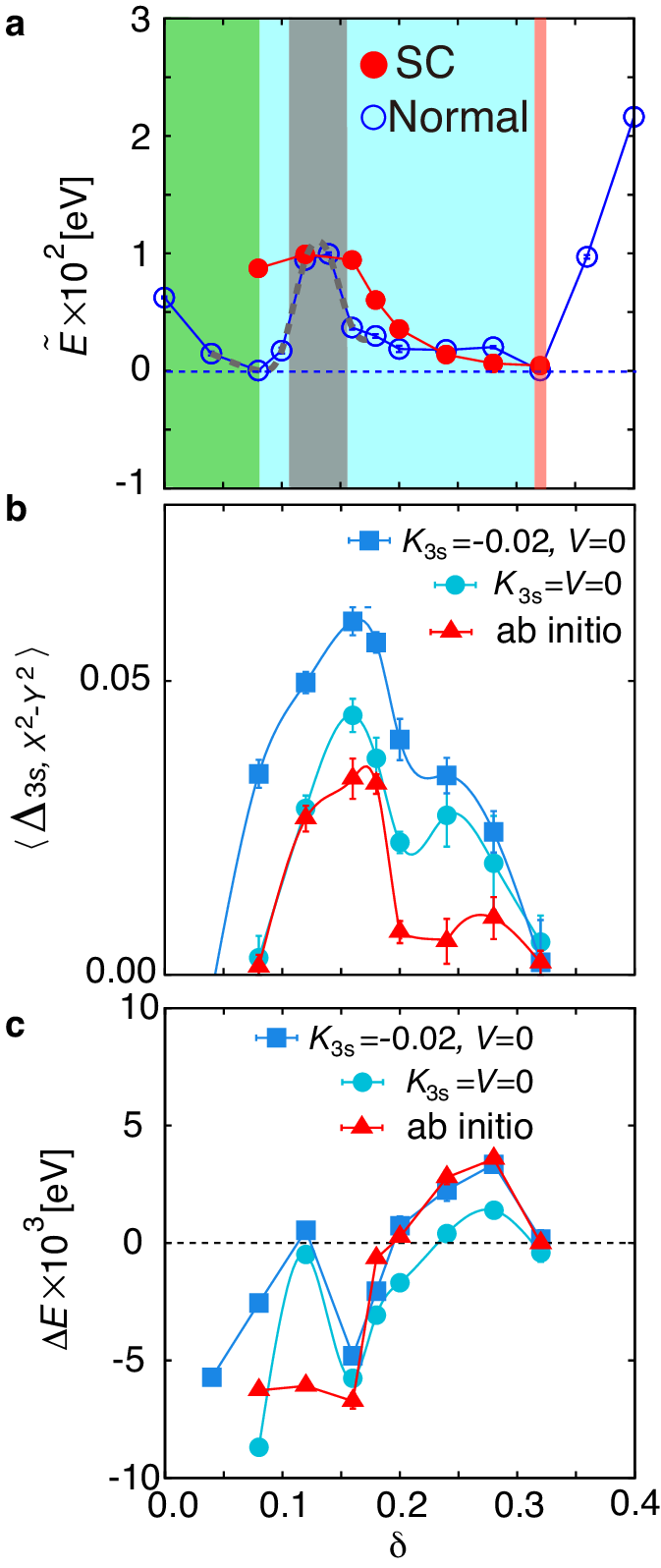}
  \end{center}
\caption{
{\bf
\tr{Doping dependence of total energies, superconducting order parameters, 
and condensation energies} by controlling off-site interactions.} 
The notations are the same as Figs.~\ref{fig:Fig1} and \ref{fig:Fig2}.
{\bf a}, $\delta$ dependence of total energy for normal ($E_{\rm n}$, open circles) and 
superconducting ($E_{\rm s}$, filled circles) phases 
for {\it ab initio} model but by switching off  off-site interactions. 
The Maxwell's construction (blue broken line) determines
the phase-separated region as  $0.08\leq\delta\leq0.32$.
{\bf b,} Comparison of 
superconducting order parameter $\langle \Delta_{3s,X^2-Y^2}\rangle$ 
among the {\it ab initio}  model (triangles), the model without off-site 
interactions (circles), and a model with an added 
attraction $K_{\rm 3s}=-0.02$eV (squares).
{\bf c,} $\delta$ dependence of the energy difference between 
the superconducting and normal states ($\Delta E=E_{\rm n}-E_{\rm s}$)
\tn{for the same cases as those in b.}
}
\label{fig:Fig5}
\end{figure}

The ground states again contain the LAF ($0<\delta<0.1$), SAF ($0.15<\delta<0.24$), and superconducting phases ($0.24<\delta<0.32$), 
as well as the spinodal ($0.1\leq\delta\leq0.15$) region {under the constraint of uniformity}, 
which are not appreciably different from the {\it ab initio} model. 
However, the phase separation region is substantially widened to $0.08\leq\delta\leq0.3$ as we see in Fig.~\ref{fig:Fig5}a.
Therefore, all of the SAF phase and most of the superconducting phase ($0.24<\delta<0.3$) become preempted by the phase separation region. 
Although 
the superconducting order parameter is substantially increased 
by switching off the off-site Coulomb interactions as we see in Fig.~\ref{fig:Fig5}b,
the stable superconducting region substantially shrinks and appears only near $\delta=0.32$ because of the 
widened phase separation region.
\tm{This result shows that the off-site Coulomb
interactions are harmful for superconductivity
in this case.} 

It was reported that the off-site interaction dramatically suppresses the superconductivity in the single-band 
Hubbard model~\cite{MisawaHubbard}, while it is not in the present case. 
The origin is that the robust first-order magnetic transition \tn{stabilized by the Hund's rule coupling cannot be suppressed by the off-site interaction here. This keeps wide area of enhanced charge fluctuations as we see later, although the phase separation itself is suppressed.}

We also add an explicit attraction, by replacing the off-site interactions to see more clearly how the pairing correlation is stabilized.
Figure~\ref{fig:Fig5}c shows that
the superconductivity is stabilized in a wider region ($0.2<\delta<0.32$) 
even at $V_{nn}=V_{nnn}=0$, when a small attraction $K_{3s}=-0.02$ eV defined in 
Methods is added, which may represent an electron-phonon interaction phenomenologically 
but rather exaggeratingly. The enhanced superconducting order parameter is also seen more clearly 
in Fig.~\ref{fig:Fig5}b in a wider region ($0.04<\delta<0.32$). These rather artificial analyses are helpful in extracting the origin of the superconductivity as detailed below. 

\trr{The crucial roles of the onsite interaction and the Hund's rule coupling in stabilizing 
the superconductivity are clear because the magnetic order itself is 
suppressed when they are weakened as clarified already~\cite{misawa2012}, 
which destroys the underlying playground of the first-order transition. 
On the other hand, as is evident in Fig.~\ref{fig:Fig5}b, 
the off-site interaction suppresses the superconducting order.  }

\begin{figure}[htb!]
  \begin{center}
    \includegraphics[width=7cm]{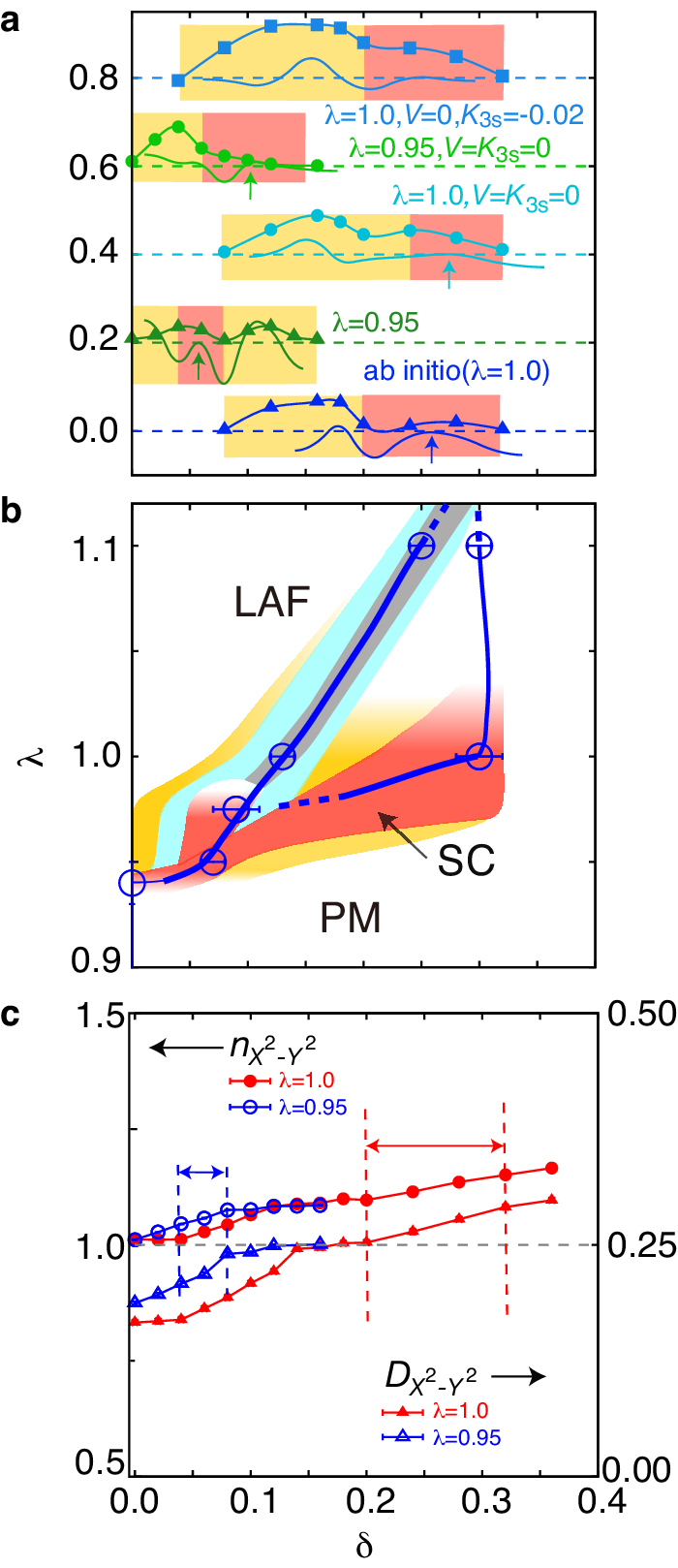}
  \end{center}
\caption{
{\bf Analysis on smoking gun for superconductivity.}~\tn{{\bf a,} 
Comparison of  \tn{$\delta$} dependence of the superconducting order
parameter \tn{$2\times\langle\Delta_{3s,X^2-Y^2}\rangle$} (symbols) 
with \tn{$0.5\times(-1/\kappa)\equiv 0.5\times(-dE_{\rm s}^2/d\delta^2)$} (curves without symbols) for various models.
The ordinates have \tn{0.2} off-set (dashed line) in sequence for clarity.
The yellow and red regions indicate nonzero $\langle \Delta_{3s,X^2-Y^2}\rangle$.
The peaks of $\langle\Delta_{3s,X^2-Y^2}\rangle$ and $-1/\kappa$ show one-to-one correspondences.
Furthermore, the stable superconducting phase (red) emerges always when $1/\kappa$ tends to vanish.
{\bf b,} The region of nonzero $\langle \Delta_{3s,X^2-Y^2}\rangle $ (either stable (red) or metastable (yellow)) surrounds the
first-order transition accompanied by the phase separation.
{\bf c,} Comparison of stable superconducting phase (two-sided arrows) with the density $n_{X^2-Y^2}$,
and the double occupation $D_{X^2-Y^2}$. 
}
}
\label{fig:Fig6}
\end{figure}

\noindent
{\bf Smoking gun for superconducting mechanism.} 
Now by collecting all the results from different parameters which provide diverse phase diagrams, 
we show that the instability toward the 
phase separation
unexceptionally enhances 
the superconducting order without ambiguity.
In Fig.~~\ref{fig:Fig6}a, $\delta$ dependence of the superconducting 
order parameter {$\langle\Delta_{3s,X^2-Y^2}\rangle$} is compared with the negative of the inverse 
charge compressibility $-1/\kappa$ defined from the second derivative of the energy with respect to $\delta$, $1/\kappa\equiv d^2E_{\rm s}/d\delta^2$
for all the cases we studied including the fully {\it ab initio}  model.
These two quantities show a good and one-to-one correspondence, 
\trr{supporting the mechanism in which} the charge fluctuation 
originally arising from the phase separation signaled 
by $1/\kappa<0$ and associated with the first-order magnetic/nematic transition   
\trr{is required for the emergence of the} superconductivity of the iron-based superconductors {LaFeAsO}.
This results is a direct evidence that the charge fluctuations induce the 
superconductivity.
\tn{Figure~{6}b further demonstrates that nonzero superconducting order emerges 
consistently around the region of the phase separation centered at first-order transition line.}

\tn{In Fig.~\ref{fig:Fig6}c, 
the region of the stable superconducting phase (two-sided arrows) is compared 
with the orbital resolved filling $n_{X^2-Y^2}$ and the double occupation 
$D_{X^2-Y^2}$ 
for two cases 
(the {\it ab initio} model ($\lambda=1$) and the case at $\lambda=0.95$). 
The stable superconducting regions are found when $n_{X^2-Y^2}$ and $D_{X^2-Y^2}$ grow fast, 
indicating the importance of the transient region in the process to reach the quasiparticle coherence for the electrons in the $X^2-Y^2$ orbital.
Although the charge fluctuation enhances the superconducting order, not only the nonzero superconducting order parameter but also 
the coherence of the  $d_{X^2-Y^2}$ orbital carrier is required to truly stabilize the superconducting phase. 
This is the reason why the stable superconductivity asymmetrically appears 
\tb {at} $\delta$ 
larger than that of the first-order transition,}
\tn{while at smaller $\lambda$ ($\lambda=0.95$), the coherence is already expected at small $\delta$, where the superconducting 
region appears symmetrically around the magnetic transition}.

\noindent
{\bf Discussion.}
Here we discuss a possible mechanism of the superconductivity \tmm{that is consistent 
with the present results indicating the one-to-one correspondence in Fig.~\ref{fig:Fig6}. 
Our result shows that the superconductivity emerges in the region where the 
strong first-order magnetic/nematic transition occurs, and the magnetic/nematic fluctuations are small. 
Thus, among
 several possible candidates of the pairing glue studied before and 
introduced in the beginning of this article, spin and orbital fluctuations are 
not supported. The only possible prominent fluctuations 
are the density fluctuations.} {When we switch off the off-site Coulomb interactions, we find that both the phase separation and
the superconducting phase are enhanced while magnetic order changes little.
This result also indicates the relevance of the density fluctuations. }

\tmm{Indeed the} superconductivity from the 
phase-separation fluctuation was proposed by
Emery, Kivelson, and Lin as the mechanism of superconductivity in
the copper oxides~\cite{EmeryKivelson}.
A general mechanism was discussed from the quantum critical 
fluctuation arising from the first-order transition of the density~\cite{Imada_PRB}.
Recent numerical calculations for the Hubbard model also suggest the 
importance of the uniform charge fluctuations in stabilizing the superconductivity~\cite{MisawaHubbard}.
In this mechanism, the instability around the spinodal decomposition 
must necessarily cause the attractive effective interaction of the carriers, because the coefficient 
of the quadratic term with respect to the density has to be negative in the energy around the spinodal point. 
It is known that the attractive interaction of the carrier is the direct cause of the pairing 
(whatever the origin of the attractive interaction is) and is very natural to induce the superconductivity, if the carriers are in the \tcy{Fermi} degeneracy region.
{One can argue that the low-energy excitation associated with the translational symmetry gives density fluctuations, which may play the role of the glue even when they are not gapless.}

However, this is not so straightforward because the superconductivity competes 
with the phase separation or other density orders. Therefore we need reliable quantitative 
calculations how the superconductivity wins or is defeated by considering 
quantum fluctuations as much as one can do. Our {\it ab initio} studies by the 
variational Monte Carlo calculation show that the superconductivity 
indeed wins in a relevant region. Our conclusion is that among possibilities, the result 
supports that the uniform charge fluctuations lead to the superconductivity observed in LaFeAs(O,F). 
For iron-based superconductors, a charge-fluctuation
mechanism was also proposed in a different context in the model with Fe 3$d$ and pnictogen (chalcogen) $p$ orbitals~\cite{ZhouKotliar}.

{ We note that the strong first-order transition by the carrier density control is of course not a sufficient condition for the realization of the superconductivity. Combination with the (orbital selective) Mottness is important, where the effective attractive interaction should take place between spin-1/2 carriers.  This is a crucial difference from other first-order transitions accompanying phase separation such as those found in perovskite manganites with the double exchange mechanism.}

Here, the uniform charge fluctuations signaled by the enhanced $-1/\kappa$ indicate the tendency for 
the aggregation of the antiferromagnetically coupled high-spin region segregated from the 
low-moment spins similarly to the case of the liquid-gas {(or binary alloy)} transition. 
The aggregation must be mediated by the attraction of the high-moment 
up and down spins (or equivalently attraction of the low-moment spins) in the stripe-type configurations. 
In the stripe configuration, 
the two {mutually 90$^{\circ}$} 
rotated configurations are degenerate and 
they interfere destructively for the nearest neighbor pair. However, the two stripe 
configurations are both constructively enhanced by the attraction of opposite spins at the 
next-nearest neighbor bonds. This attraction causes the $s\pm$ Cooper pair as well. Namely, the phase segregation and the singlet 
pairing with the $s\pm$ symmetry are two-sides of coins and are simultaneously 
driven by the next-neighbor attraction of the pair caused 
by the high to low-spin transition and the underlying rapid Mottness crossover.

{The mechanism we revealed may be experimentally tested by measuring the superconductivity at the interface of
two bulk superconductors with different doping concentrations, one being the mother material with 
antiferromagnetic order and the other being the overdoped non superconducting materials as in the experiment 
for the interface of the cuprate superconductors~\cite{WuBozovic}. 
When the average electron density of the 
interface is tuned around the phase separation region, we predict that the 
transition temperature is kept constant at the optimum value~\cite{MisawaHubbard}.}
In fact, in the thin film of FeSe on SrTiO$_{3}$,
\tr{high-$T_{\rm c}$} superconductivity ($T_{\rm c}\gtrsim 50$K) 
was reported~\cite{Qing_2012}.
{The phase separation in \tmmm{$A_x$Fe$_{2-y}$Se$_2$, ($A$=K, Rb)}~\cite{Texier_2012,Li_2012,Yu_2013,Wang_2014} 
into antiferromagnetic and superconducting regions with a high critical 
temperature above 60 K was also reported in experiments. These are also consistent 
with the present results, because the optimum doping concentration 
can be realized near the interface of the two phases.
The systematic study on the electron 
concentration dependence at the interface will clarify more clearly the mechanism revealed here.}

{We have not discussed the role of magnetic fluctuations in detail.
{This is because, around the strong first-order transitions, the magnetic fluctuations are not
significant as we already discussed.}
Near the
magnetic quantum critical point around $\lambda=0.94$ seen in Fig. 4, however,
the magnetic fluctuations are expected to 
{contribute}, while the density fluctuations are suppressed because of the conversion of the first-order
transition to a continuous one. Indeed the functional renormalization group
study combined with the first-principles approach similar to ours suggests the
role of magnetic fluctuations~\cite{PlattPRB2011}. The functional
renormalization group is a complementary weak-coupling approach, where weak
instability to the superconductivity 
\tb{can be easily} studied. 
The mVMC is able to 
study \tb{more easily} the strong coupling superconductivity.
We\tb{, however,}
see in Fig. 4 that the superconducting phase is extended \tb{even}
to the region around the quantum critical point, \tb{implying the applicability of
mVMC in both strong and weak coupling regions.}
\tb{In most of the phase diagram, the density fluctuations play the dominant role in
stabilizing \tr{high-$T_{\rm c}$} superconductivity as the strong coupling mechanism.}
} 

\tgf{More quantitative {\it ab initio} analyses including other 
families of the iron-based superconductors and the role of phonons 
as well as guiding principle for raising $T_{\rm c}$  
are intriguing and challenging issues left for future studies.  
}

\clearpage
\noindent
{\bf Methods.} \\
{{\bf Details of model Hamiltonian.}}
In the present approach, we seek for an undoubted way of extracting the origin and
mechanism of the experimentally observed superconductivity in the
iron-based superconductors based on material-dependent realistic calculation.
For this purpose, we employ the {\it ab initio} way of deriving the effective
Hamiltonian without any adjustable parameters. Then we solve the effective model
as accurate as possible within the available methods.

{The Hamiltonian of low-energy effective model for a two-dimensional layer of LaFeAsO is given by}
\begin{align}
\mathcal{H} &= \mathcal{H}_{0}+\mathcal{H}_{\rm int}  \label{Eq:Ham1_M}\\
\mathcal{H}_{\rm int} &= \mathcal{H}_{\text{\rm on-site}}+\mathcal{H}_{\text{\rm off-site}}  \label{Eq:Ham1_b_M}\\
\mathcal{H}_{0}&=\sum_{\sigma} \sum_{i,j} \sum_{\nu,\mu}  
  t_{i,j,\nu,\mu} 
                   c_{i,\nu,\sigma}^{\dagger} 
                   c_{j,\mu,\sigma}    \label{Eq:Ham2_M}\\
\mathcal{H}_{\text{\rm on-site}}&= \frac{1}{2} \sum_{\sigma, \sigma'} \sum_{i} \sum_{\nu,\mu} 
  \biggl\{ U_{i,i,\mu,\nu} 
                   c_{i,\nu,\sigma}^{\dagger} 
                   c_{i,\mu,\sigma'}^{ \dagger}
                   c_{i,\mu,\sigma'} 
                   c_{i\nu,\sigma}  \notag \\ 
&+ J_{i,i,\mu,\nu } 
\bigl(c_{i,\nu,\sigma}^{\dagger} 
      c_{i,\mu,\sigma'}^{ \dagger}
      c_{i,\nu,\sigma'} 
      c_{i,\mu,\sigma}  \notag \\
   &+c_{i,\nu,\sigma}^{\dagger} 
      c_{i,\nu,\sigma'}^{\dagger}
      c_{i,\mu,\sigma'} 
      c_{i,\mu,\sigma}\bigr) \biggr\},\label{Eq:Ham_on_M} \\
\mathcal{H}_{\text{\rm off-site}} &=V_{nn}\sum_{\langle i,j\rangle,\nu,\mu} n_{i\nu}n_{j\mu}
+V_{nnn}\sum_{\langle\langle k,l\rangle\rangle,\nu,\mu} n_{k\nu}n_{l\mu}
\label{Eq:Ham_off_M}
\end{align}
Here, $t_{i,j,\nu,\mu}$ contains single-particle levels and transfer integrals, while $U_{i,i,\nu,\mu}$ and
$J_{i,i,\nu,\mu}$ are screened Coulomb and exchange
interactions, respectively. The exchange interaction $J_{i,i,\nu,\mu}$ consists of 
the Hund's rule coupling (conventionally denoted as 
$J_{H\nu\mu}\equiv J_{i,i,\nu,\mu}$) in the first term and the pair hopping in the second term.
We use the transfer integrals up to the fifth neighbors~\cite{miyake2010}, which well reproduce the
LDA band structures.
In the off-site Coulomb interactions $H_{\text{off-site}}$, 
$V_{nn}$ ($V_{nnn}$) represents the nearest-neighbor (next-nearest-neighbor) Coulomb interactions and
$n_{i\nu}$ denotes the orbital occupation of $\nu$th orbital at the $i$th site~\cite{miyake2010}.
From the $ab$ $initio$ downfolding procedure, we estimate $V_{nn}=0.4$ eV and $V_{nnn}=0.2$ eV~\cite{nakamura2010}.
{Further neighbor interactions are exponentially small because the two-dimensional effective model takes into
account metallic screening from other layers.}
Because the off-site interactions do not 
appreciably depend on the combinations of orbitals and are distributed within 0.01 eV, 
we ignore the orbital dependence.
Other off-site interactions such as the off-site direct exchange interactions 
are also less than 0.01 eV on average
and we also ignore them.

\tn{All of the model parameters were derived in the so-called downfolding procedure~\cite{miyake2010}. In this procedure, the global electronic structure is calculated by the {\it ab initio} density functional calculations. Then the degrees of freedom whose energies are located far away from the Fermi level are traced out,  leaving the {\it ab initio} effective model appropriate near the Fermi level, namely, for Fe $3d$ five orbitals on the experimental crystal structure. Then the {\it ab initio} two-dimensional model for a layer, where Fe atoms are arrayed on a square lattice, is derived after the dimensional downfolding~\cite{nakamura2010}.  The detailed description of the Hamiltonian parameters are found in refs.~\cite{miyake2010,nakamura2010,misawa2012}.}

\tn{The doping concentration ($\delta \equiv 1-N_e/N_s$) dependence is studied 
in this article by changing the electron number $N_e$ in a layer containing $N_s$ iron sites 
in the periodic boundary condition, where other parameters in the Hamiltonian are 
assumed to be fixed through the doping process. Though it is adequate within 
the present work, in the main text, we also discuss possible 
modification of the Hamiltonian parameters by heavy doping.}\\

\noindent
{{\bf Details of mVMC.}}
We study the possibility of superconductivity 
in the model (\ref{Eq:Ham1_M})-(\ref{Eq:Ham_off_M}) by a many-variable variational Monte Carlo (mVMC) method formulated in the literature~\cite{TaharaVMC_Full}. 
In the mVMC calculations, we study the ground state properties by 
employing a generalized \tr{BCS} type wave function with the 
quantum number projection and the Gutzwiller\cite{Gutzwiller} and Jastrow factors~\cite{Jastrow}; 
$|\psi\ra = \sP_{\rm G}\sP_{\rm J}\sL^{S=0}|\phi_{\rm pair}\ra$. 
Here, $\sL^{S=0}$ is the spin projection operator to the total spin $S=0$ subspace; 
$\sP_{\rm G}$ and $\sP_{\rm J}$ are the Gutzwiller and Jastrow factors, respectively~\cite{TaharaVMC_Full}.
The spin projection is performed onto the $S=0$ singlet subspace.
The Gutzwiller factor punishes the double occupation 
of electrons by 
$\sP_{\text{G}} = \exp(-\sum_{i,\nu} g_{i\nu} n_{i\nu \uparrow} n_{i\nu \downarrow})$ 
where $n_{i\nu\sigma} = c_{i\nu \sigma}^\dagger c_{i\nu \sigma}$.
The Jastrow factor is introduced up to the next-nearest-neighbor sites as 
$\sP_{\text{J}} = \exp(-\frac{1}{2} \sum_{i,j} v_{ij\nu\mu} n_{i\nu} n_{j\nu})$,
where $n_{i\nu} = \sum_\sigma n_{i\nu \sigma}$. 
The one-body part $|\phi_{\rm pair}\ra$ is the generalized pairing wave function defined as
$|\phi_{\rm pair}\rangle=(\sum_{\nu,\mu=1}^{5}\sum_{i,j=1}^{N_{\rm s}}f_{ij\nu\mu}c_{i\nu\uparrow}^\dag c_{i\mu\downarrow}^\dag)^{N_e/2} |0 \rangle$, 
where $N_e$ is the number of electrons. 
In this study, we restrict the variational parameters, $g_{i\nu}$, $v_{ij\nu\mu}$ to have a $2\times 1$ structure, 
and $f_{ij\nu\mu}$ to have a $2\times2$ sublattice structure. 
The number of variational parameters are 10 for $g_{i\nu}$, 220 for $v_{ij\nu\mu}$, and
\tn{100$N_s$}  for $f_{ij\nu\mu}$.
All the variational parameters are simultaneously 
optimized by using the stochastic reconfiguration method~\cite{Sorella_PRB2001,TaharaVMC_Full}.
Our variational wave function $|\psi\ra$ can flexibly describe superconducting, 
\tr{AF}, and \tr{PM}  phases as well as their fluctuations 
on {an} equal footing.
The calculations were done up to $10\times 10$ sites.

Monte Carlo sampling of real space configurations of the electrons is
employed to calculate physical quantities following
the standard procedure~\cite{TaharaVMC_Full}.
\tm{The acceptance ratio of the Monte Carlo sampling
is typically more than 10\%.}
\tm{Here, we define the autocorrelation as 
\begin{equation*}
A(t)=\frac{1}{N_{\rm s}}\sum_{i,\sigma}n_{\sigma}(r_{i},t)n_{\sigma}(r_{i},0)-2\bar{n}^2,
\end{equation*}
where $n_\sigma(r_{i},t)$ the number of particle at $i$th site, and $t$ represents
 Monte Carlo step.
$\bar{n}$ is the averaged density per spin 
\tr{, which is defined as} $\bar{n}=N_{e}/2N_{s}$. 
The autocorrelation vanishes within
10-20 Monte Carlo steps.
We show an example of $A(t)$ in Supplementary \tr{Fig.~3.}}
The number of Monte Carlo samples for the calculation of
physical quantities 
is typically 128 000.
The statistical error of the Monte Carlo sampling is estimated from
a number of independent bins typically around five.

\noindent
{{\bf Details of physical properties.}}
To reveal physical properties and determine the phase diagram of the {\it ab initio} model,
we calculated \tn{orbital resolved filling $n_{\nu}$, orbital dependent double occupation $D_{\nu}$,} equal-time spin structure factors $m({\bm q})^2$, its orbital-diagonal component $m_{\nu}({\bm q})^2$, 
the equal-time superconducting correlation $P_{\alpha}({\bm r})$, \tn{and its orbital resolved component $P_{\alpha,\nu,\mu}({\bm r})$}, which 
are defined as
\tn{
\begin{align}
n_{\nu}&= \frac{1}{N_s}\sum_{\sigma,i} \langle n_{i,\nu,\sigma}\rangle \\
D_{\nu}&= \frac{1}{N_s}\sum_{\sigma,i} \langle n_{i,\nu,\uparrow}n_{i,\nu,\downarrow}\rangle \\
m(\bm{q})^{2}&= \sum_{\nu,\mu}m_{\nu,\mu}(\bm{q})^{2},\\
m_{\nu}(\bm{q})^{2}&= \sum_{\nu}m_{\nu,\nu}(\bm{q})^{2},\\
m_{\nu,\mu}(\bm{q})^{2}&= {\frac{4}{3N_{\rm s}^2} \sum_{i,j}\langle\bm{S}_{i\nu}\cdot\bm{S}_{j\mu}\rangle e^{i\bm{q(r_{i}-r_{j})}}} ,\\
\vec{S}_{i\nu}&=\frac{1}{2}\sum_{\sigma,\sigma^{\prime}}c_{i\nu,\sigma}^{\dagger}\vec{\sigma}_{\sigma\sigma^{\prime}}c_{i\nu,\sigma^{\prime}},\\
P_{\alpha}(\bm{r})&=\sum_{\nu,mu,\xi,\eta}
P_{\alpha,\nu,\mu,\xi,\eta}(\bm{r}), \\
P_{\alpha,\nu,\mu}(\bm{r})&=
P_{\alpha,\nu,\mu,\mu,\nu}(\bm{r}), \\
P_{\alpha,\nu,\mu,\xi,\eta}(\bm{r})&=\frac{1}{2N_{\rm s}}\sum_{\bm{r}_{i}}
[\langle\Delta_{\alpha,\nu,\mu}^{\dag}(\bm{r}_{i})\Delta_{\alpha,\xi,\eta}(\bm{r}_i+\bm{r})\rangle+ {\rm H.c.}],
\end{align}
}
where $\vec{\sigma}$ represents Pauli matrix.
{In actual calculations, to reduce numerical cost, we restrict the summation with 
respect to $\bm{r}_{i}$ within $2\times 2$ sublattice.}
The magnetic order parameter is estimated after the size extrapolation of the finite size data from $4\times 4$ to $10\times 10$ to the 
thermodynamic limit.  The extrapolation is performed as a linear fitting of $m(\bm{Q})$, as a function of the inverse linear dimension $1/L$, where $\bm{Q}$ is the peak position of $m(\bm{q})$.  Stripe order is determined from the Bragg peak at $(\pi,0)$.

Superconducting order parameter and its orbital diagonal component are defined as
\tn{
\begin{align*}
\Delta_{\alpha,\nu,\mu}(\bm{r}_i)&=\frac{1}{\sqrt{2}}
\sum_{\bm{r}}f_{\alpha}(\bm{r})({c}_{\bm{r}_i,\nu\uparrow}{c}_{\bm{r}_i+\bm{r}\mu\downarrow}-{c}_{\bm{r}_i\mu\downarrow}{c}_{\bm{r}_i+\bm{r}\nu\uparrow}),
\\
\Delta_{\alpha,\nu}(\bm{r}_i)&=\frac{1}{\sqrt{2}}
\sum_{\bm{r}}f_{\alpha}(\bm{r})({c}_{\bm{r}_i,\nu\uparrow}{c}_{\bm{r}_i+\bm{r}\nu\downarrow}-{c}_{\bm{r}_i\nu\downarrow}{c}_{\bm{r}_i+\bm{r}\nu\uparrow}).
\end{align*}
}
Here, $f_{\alpha}(\bm{r})$ is the form factor that
describes the symmetry of the superconductivity.
In the present work, we examined the four {possible} symmetries of the superconducting states; 
$\alpha=2s{({\rm gappless}~s_{\pm})},
~2d{(d_{x^{2}-y^{2}})},
~3s{({\rm gapped}~s_{\pm})}$, and
$3d~{(d_{xy})}$, whose form factors are defined as
\begin{align*}
f_{2s}(\bm{r})&=\delta_{r_{y},0}(\delta_{r_{x},1}+\delta_{r_{x},-1})-\delta_{r_{x},0}(\delta_{r_{y},1}+\delta_{r_{y},-1}) ,  \\
f_{2d}(\bm{r})&=\delta_{r_{y},0}(\delta_{r_{x},1}+\delta_{r_{x},-1})+\delta_{r_{x},0}(\delta_{r_{y},1}+\delta_{r_{y},-1}) ,  \\
f_{3s}(\bm{r})&=(\delta_{r_{x},1}+\delta_{r_{x},-1})(\delta_{r_{y},1}+\delta_{r_{y},-1}),\\
f_{3d}(\bm{r})&=(\delta_{r_{x},1}-\delta_{r_{x},-1})(\delta_{r_{y},1}-\delta_{r_{y},-1}),
\end{align*}
where $\delta_{i,j}$ denotes the Kronecker's delta and $\bm{r}=(r_{x},r_{y})$.

We mainly study the case of $f_{\alpha}$ with $\alpha=3s$ often called $s\pm$  
symmetry, where the pairing is between the electrons on the next-nearest neighbor sites,
because this symmetry of the pairing is the only one that survives in the realistic model. 
The nodal struture of this pairing is shown in 
{Supplementary} \tr{Fig.~4a}. 
{Initial conditions with other symmetries converge to the normal state.}

The order 
is determined from the long-ranged level-off part of the averaged pairing correlation 
$\langle {\Delta}_{3s,\nu}\rangle =\sqrt{\lim_{\bm{r}\rightarrow \infty}P_{3s,\nu,\nu}({\bm r})}$,
\tn{where we omit the contribution from the orbital off-diagonal pairing, because they are expected to be smaller.} 

In {Supplementary} \tr{Fig.~4b}, we plot the 
superconducting correlation $P_{3s,\nu}(\vec{r})$ for an 
example of $\nu=X^{2}-Y^{2}$ as a function of the distance for the 
superconducting phase for the {\it ab initio} model at $\delta=0.24$. 
In the normal phase, the superconducting correlation shows 
a power-law decay ($\sim r^{-3}$),
and its amplitude is comparable to that of the non-interacting case.
In contrast to this, 
in the superconducting phase,
the superconducting correlation is saturated to a nonzero constant value as in Fig.~\ref{fig:Fig1}b.

To see whether the superconducting correlation 
is saturated to a nonzero value,
we define long-range average of the superconducting correlation as
\begin{align*}
\bar{P}_{\alpha,\nu}&=\frac{1}{M}\sum_{R<r=|\bm{r}|\leq\sqrt{2}L}P_{{\alpha},\nu}(\bm{r}), \\
\bar{\Delta}_{\alpha,\nu}&=\sqrt{\bar{P}_{\alpha,\nu}}
\end{align*}
where $M$ is the number of vectors satisfying $R<r\leq\sqrt{2}L$ ($L$ is a linear dimension of the system size).
For the present purpose, $R=3$ is practically a sufficient criterion to see whether 
the pairing order-parameter correlation is saturated to a nonzero value and 
$\bar{\Delta}_{\alpha,\nu}$ offers a good measure for the order parameter in the long-range ordered superconducting state.
To further reduce the finite size effect, we subtract $\bar{\Delta}_{\rm Normal}$ calculated for the 
normal state from that for the superconducting state $\bar{\Delta}_{\rm SC}$. 
In the actual calculations $\bar{\Delta}$ in the normal state is nonzero 
because of the finite size effects. In the analyses we use $\langle \Delta \rangle=\bar{\Delta}_{\rm SC} -\bar{\Delta}_{\rm Normal}$. 
\tn{We observed that $\langle \Delta_{3s,X^2-Y^2
} \rangle$ is always dominant over other orbital contributions.}

\noindent
{{\bf Details of attractive interactions.}}
When we examine the effects of attractive interactions, we add an attractive interaction term, defined as
\begin{align}
K_{3s}&\sum_{i,\nu}\Big[\Delta_{3s,\nu}^{\dagger}(i)\Delta_{3s,\nu}(i)+\Delta_{3s,\nu}(i)\Delta_{3s,\nu}^{\dagger}(i)\Big],
\label{K3s}
\end{align}
where $\Delta_{3s,\nu}(i)$ represents
the superconducting order parameter of $\nu$th orbital at $i$th site for
the gapped $s_{\pm}$-wave superconductivity.
Similar interactions can be derived by considering the 
phonon degrees of freedom~\cite{Hirsch1987}. 
In actual calculations, we {dropped} the one-body part that originates
from the commutation relation. We also {dropped} the AF interactions
term that is proportional to $\vec{S}_{i}\cdot \vec{S}_{j}$ and contained in Eq.(\ref{K3s}), which
induces the AF order. \\

\noindent
{{\bf Details of phase diagram.}}
\tn{The phases in the phase diagram are determined by the lowest energy state
if more than one locally stable states are found. 
The stripe type magnetic order with the large (small) ordered moment is 
shown as dark (light) green area and the superconducting phase is shown as 
red area in the figures of the main text.  In addition, thermodynamically prohibited region 
by the phase separation (phase separation region) (illustrated as light blue areas in the figures) 
is determined by the Maxwell's construction, where the region of the 
ground-state energies above a common tangent of the two points in the $\delta$ 
dependence of the ground-state energy is identified as the phase separation region. 
If the second derivative of the energy $d^2E/d\delta^2$ is negative inside the 
phase separation region, the region is locally unstable to inhomogeneity 
even under an infinitesimal perturbation and called spinodal 
region (gray area in the figures). The phase separation and accompanied 
spinodal region are natural consequences of the first-order magnetic transitions. \\}   
{Strictly speaking, the stable uniform states are prohibited in 
the spinodal region in the thermodynamic limit.
Precise estimate of the spinodal region from the calculation 
of finite-size systems is difficult, because the negative curveture of the energy is eventually prohibited in the thermodynamic limit.}

\vspace{\baselineskip}
\noindent
{\bf Acknowledgements}
The authors thank
Daisuke Tahara and Satoshi Morita
for providing them with efficient mVMC codes.
{They also thank Kazuma Nakamura and Takashi Miyake for providing them
parameters of $ab$ $initio$ models.}
This work is financially supported by MEXT HPCI Strategic Programs
for Innovative Research (SPIRE) and Computational Materials Science Initiative (CMSI).
Numerical calculation was partly carried out at the Supercomputer Center,
Institute for Solid State Physics, Univ. of Tokyo.
Numerical calculation was also partly carried out at K computer at
RIKEN Advanced Institute for Computational Science (AICS)
{under grant number hp120043, hp120283 and hp130007}.
This work was also supported by Grant-in-Aid for
Scientific Research {(No. 22104010, No. 22340090, and No. 23740261)} from MEXT, Japan.\\

\noindent
{\bf Author Contributions} M.I. and T.M. designed the project and directed the investigation. T.M.
performed the simulations and prepared the figures. Results were analyzed and the paper was written by all authors. \\

\noindent
{\bf Supplementary Information} is linked to the online version of the paper. \\

\noindent
{\bf Competing financial interests:} The authors declare no competing financial interests.

\clearpage
{\bf SUPPLEMENTARY INFORMATION.}
\renewcommand{\figurename}{{\bf Supplementary Figure}}
\setcounter{figure}{0}
\begin{figure}[h!]
  \begin{center}
    \includegraphics[width=7cm]{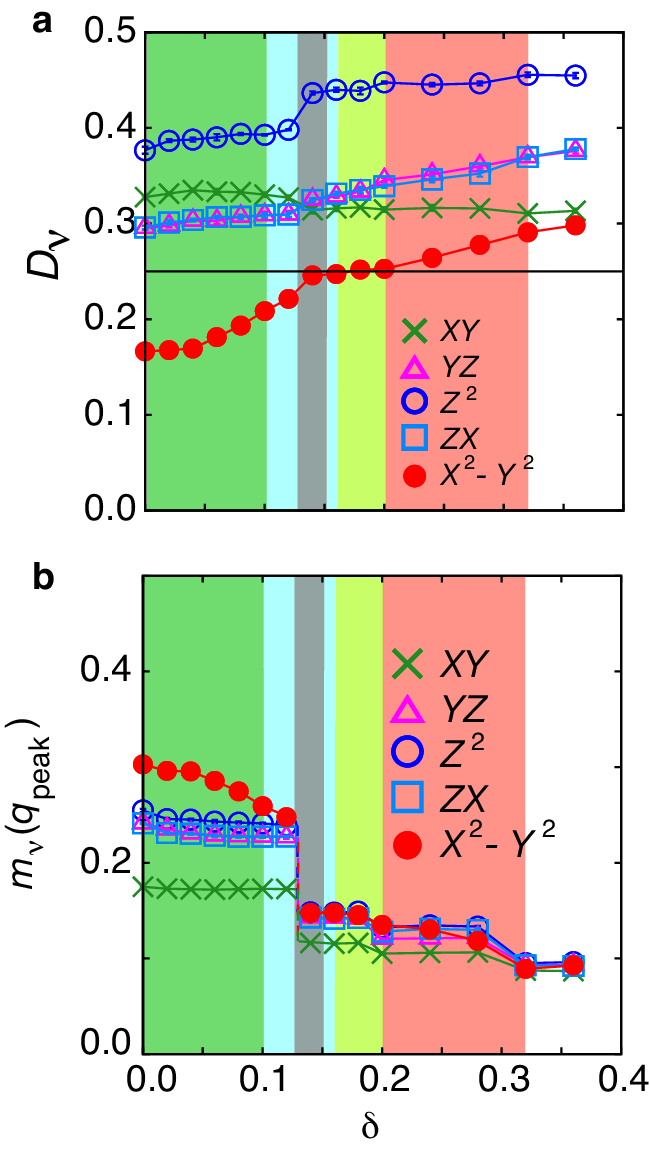}
  \end{center}\caption{
{\bf a,~}Orbital resolved diagonal components
of stripe-type magnetic ordered moment
$m_{\nu}\equiv m_{\nu,\nu}({\bm q}(\pi,0))$ at the peak momentum $q_{\rm peak}=(\pi,0)$
for the {\it ab initio} model
at the largest available system size ($10 \times 10$).
The notation for the colored area is the same as Fig. 1.
{\bf b,}~Orbital resolved double occupancies for the {\it ab initio} model
at the largest available system size ($10 \times 10$).
}
\label{fig:FigS2}
\end{figure}%

\begin{figure}[b!]
  \begin{center}
    \includegraphics[width=7cm]{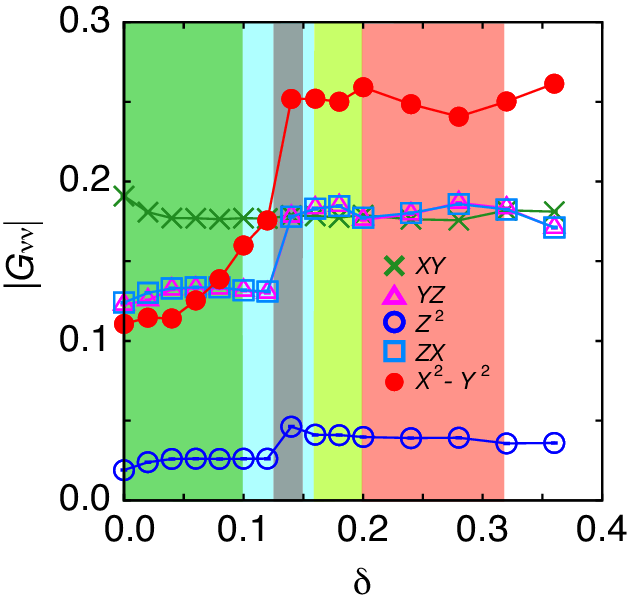}
  \end{center}\caption{
Doping dependence of absolute values of nearest neighbor
diagonal Green functions 
$|G_{\nu\nu}|=|\langle c_{i\uparrow}^{\dagger}c_{j\uparrow}
+c_{i\downarrow}^{\dagger}c_{j\downarrow}\rangle|$, where
$i$, $j$ are nearest sites.
We take $N_{s}=10\times 10$ and $\lambda=1$~($ab$ $initio$ model).
}
\label{fig:FigS4}
\end{figure}%

\begin{figure}[b!]
  \begin{center}
    \includegraphics[width=7cm]{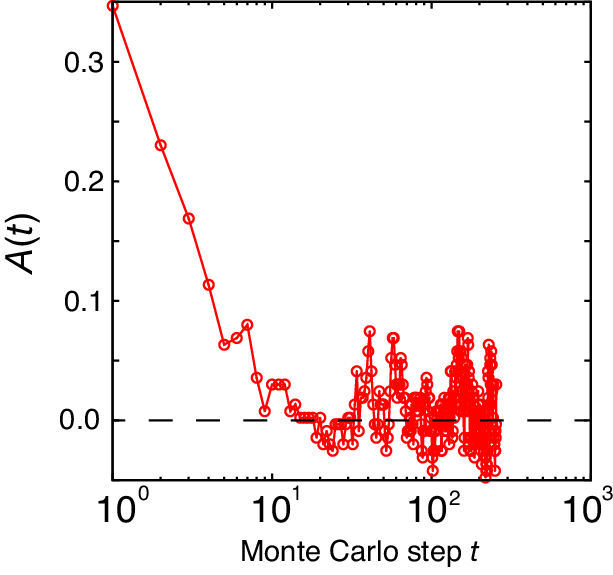}
  \end{center}\caption{
Autocorrelation as a function Monte Carlo steps.
We take $N_{s}=6\times 6$, $\delta=0$, 
and $\lambda=1$~($ab$ $initio$ model).
}
\label{fig:FigS3}
\end{figure}%

\begin{figure}[h!]
  \begin{center}
   \includegraphics[width=8cm,clip]{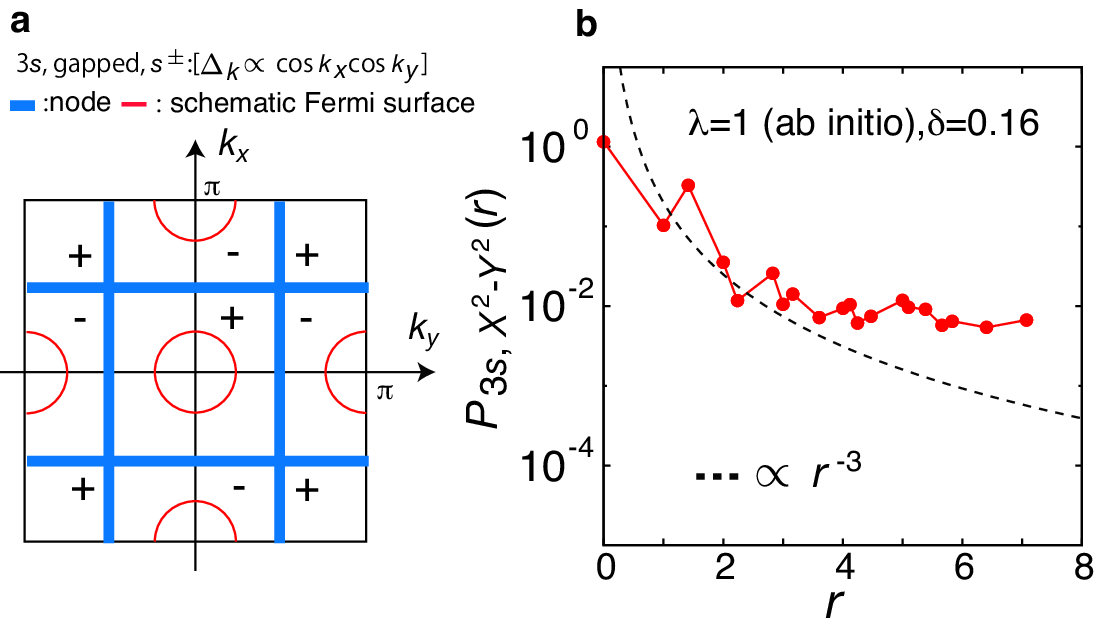}   
  \end{center}
\caption{
{\bf a,} Schematic illustration of superconducting gap structure for $3s$ (gapped $s_{\pm}$)
symmetry (with nodes illustrated by blue thick lines). We also plot the schematic
Fermi surface of LDA band structures for LaFeAsO (red thin circles).
{\bf b,} Superconducting correlations as function of distance $r$ at $\delta=0.24$ and $\lambda=1.0$@{({\it ab initio} model)}
for superconducting {ground state} (red close circles). 
System size is {$N_{\rm s}=L\times L$ with $L=10$}.
The short-ranged part follows the decay proportional to $r^{-3}$ while it levels off at a nonzero value at
long distances (roughly at $r\geq3$ {in this case}).
For comparison, we plot the asymptotic $r^{-3}$ behavior 
(broken line) of the superconducting correlations for 
the non-interacting systems, 
which is expected for two-dimensional metals.}
\label{fig:FigS1}
\end{figure}%

\clearpage

\renewcommand{\tablename}{{\bf Supplementary Table}}
\begin{table*}
\caption{
Effective on-site Coulomb ($U$)/exchange ($J$) interactions between two electrons
on the same iron site in the $ab$ $initio$ model for LaFeAsO (in eV).
}
\label{tab:Parameters}
{\scriptsize 
\begin{tabular}{ccccccccccccccc}
  \\
\hline  \\
LaFeAsO  &      &      & $U$   &      &          &  &     &     &      &  $J$   &      &    \\
\hline \\
        & $XY$ & $YZ$ & $Z^2$ & $ZX$ & $X^2-Y^2$ & &     & $XY$ & $YZ$ & $Z^2$ & $ZX$ & $X^2-Y^2$  \\
\hline
$XY$ & 2.62  & 1.39  & 1.37 & 1.39 & 1.50 &   & $XY$     &      & 0.46 & 0.57 & 0.46 & 0.23 \\
$YZ$ & 1.39  & 2.02 & 1.56 & 1.21 & 1.11  &   & $YZ$     & 0.46 &      & 0.33 & 0.37 & 0.35 \\
$Z^2$ & 1.37  & 1.56 & 2.43 & 1.56 & 1.10 &   & $Z^2$    & 0.57 & 0.33 &      & 0.33 & 0.42 \\
$ZX$ & 1.39  & 1.21 & 1.56 & 2.02 & 1.11  &   & $ZX$     & 0.46 & 0.37 & 0.33 &      & 0.35 \\
$X^2-Y^2$ & 1.50  & 1.11 & 1.10 & 1.11 & 1.50 & &$X^2-Y^2$ & 0.23 & 0.35 & 0.42 & 0.35 &      \\
\hline \\
\end{tabular}
} 
\end{table*}

\begin{table*}[htb] 
\caption{Transfer integrals in the $ab$ $initio$ model 
for LaFeAsO, $t_{\mu\nu}(R_X, R_Y, R_Z)$, 
where $\mu$ and $\nu$ specify symmetry of $d$ orbitals; 
1 for ${XY}$, 
2 for ${YZ}$, 
3 for ${Z^{2}}$, 
4 for ${ZX}$, 
and 5 for ${X^2-Y^2}$ orbitals. 
Symmetry operations of 
$\sigma_{Y}$, $I$, and $\sigma_{d}$ 
change $t_{\mu\nu}(R_X, R_Y, R_Z)$ to $t_{\mu\nu}(R_X, -R_Y, R_Z)$, $t_{\mu\nu}(-R_X,-R_Y,R_Z)$, 
and $t_{\mu\nu}(R_Y,R_X,R_Z)$. Notice also that $t_{\mu\nu}(\bR)=t_{\nu\mu}(-\bR)$. 
Units are given in meV. 
} 
\
{\scriptsize 
\begin{tabular}{c|rrrrrrrrrrc} 
\hline  \\ [-4pt]
LaFeAsO \\ [+2pt] 
\hline \\ [-4pt]
\backslashbox{$(\mu, \nu)$}{$\bR$} 
& \big[0,0,0\big] 
& \big[$\frac{1}{2}$,$-\frac{1}{2}$,0\big] 
& \big[1,0,0\big] 
& \big[1,$-$1,0\big] 
& \big[$\frac{3}{2}$,$-\frac{1}{2}$,0\big] 
& \big[2,0,0\big] 
& \big[0,0,$\frac{c}{a}$\big] 
& \big[$\frac{1}{2}, -\frac{1}{2}, \frac{c}{a}$\big] 
& $\sigma_{Y}$ 
& $I$ 
& $\sigma_{d}$ \\ [+4pt]
\hline \\ [-8pt]
$(1,1)$&790 & $-$315 &$-$67 & $-$19& $-$2 &  1 & $-$2&1 & +  &  +  &   + \\
$(1,2)$&  0 & 253 &138  & 1  & 10 &  0   &  0  &   0 & +  & $-$ & $-$(1,4)  \\
$(1,3)$&  0 & $-$301 & 0  &1  & $-$18 &  0   &  0  &   0 & $-$ &  +  &   +   \\
$(1,4)$&  0 & 253 & 0  & 1  &33 &  0   &  0  &   $-$1 & $-$ & $-$ &$-$(1,2) \\
$(1,5)$&  0 & 0 & 0  &    0  & 10 &  0&  0  &   $-$2 & $-$ &  +  &    $-$    \\
$(2,2)$& 1099 & 206 &135 & 12  & 9 &  5   &  1  &   7 &  +  &  +  & (4,4)   \\
$(2,3)$&  0 & $-$73 &0  & $-$2& $-$1 &  0 &  0  &  2 & $-$ & $-$ &$-$(4,3) \\  
$(2,4)$&  0 & 137 & 0  & $-$18& $-$9 &  0   &  0  &   1 & $-$ &  +  & (4,2) \\
$(2,5)$&  0 & 165 &   0  & -4  & 10 &  0   &  0  &   3 & $-$ & $-$ & (4,5)  \\
$(3,3)$&890 &72 & $-$13& $-$38  & $-$15 & $-$18 &$-$6 &$-$2 &  +  &  +  &   +\\
$(3,4)$&  0 & 73 & 137 & 2  & $-$3 &  0 &  0  & $-$1 &  +  & $-$ & $-$(3,2)  \\
$(3,5)$&  0 & 0 & $-$159 & 0  & 1 &  17 &  3  &   $-$3 &  +  &  +  &   $-$   \\
$(4,4)$& 1099 &206 & 345& 12  & 36 &70 &  1  &   0 &  +  &  +  &  (2,2)  \\
$(4,5)$&  0 & $-$165 & 19 & 4 & $-$11 &  0 &  0  &   1 &  +  & $-$ & (2,5) \\
$(5,5)$& 1255 & $-$152 &118& $-$24  & 30 & $-$28 &  1 &$-$2 &  +  &  +  & + \\
\hline 
\end{tabular}
}
\label{tab:Transfers} 
\end{table*}


\begin{thebibliography}{99}
\expandafter\ifx\csname natexlab\endcsname\relax\def\natexlab#1{#1}\fi
\expandafter\ifx\csname bibnamefont\endcsname\relax
  \def\bibnamefont#1{#1}\fi
\expandafter\ifx\csname bibfnamefont\endcsname\relax
  \def\bibfnamefont#1{#1}\fi
\expandafter\ifx\csname citenamefont\endcsname\relax
  \def\citenamefont#1{#1}\fi
\expandafter\ifx\csname url\endcsname\relax
  \def\url#1{\texttt{#1}}\fi
\expandafter\ifx\csname urlprefix\endcsname\relax\def\urlprefix{URL }\fi
\providecommand{\bibinfo}[2]{#2}
\providecommand{\eprint}[2][]{\url{#2}}

\bibitem[{\citenamefont{Kamihara et~al.}(2008)\citenamefont{Kamihara, Watanabe,
  Hirano, abd Hosono}}]{kamihara2008}
\bibnamefont{Kamihara,}\bibinfo{author}{\bibfnamefont{~Y.,}
  }\bibnamefont{Watanabe,}\bibinfo{author}{\bibfnamefont{~T.,}
  }\bibnamefont{Hirano,}\bibinfo{author}{\bibfnamefont{~M.,}
  }\bibnamefont{\&} \bibnamefont{Hosono,}\bibinfo{author}{\bibfnamefont{~H.}
  } Iron-based layered superconductor La[O$_{1-x}$F$_{x}$]FeAs~($x=0.05$-$0.12$) with $T_{\rm c}= 26$K.
  \bibinfo{journal}{{\it J. Am. Chem. Soc.}} \textbf{\bibinfo{volume}{130}},
  \bibinfo{pages}{3296--3297} (\bibinfo{year}{2008}).

\bibitem[{\citenamefont{Kuroki et~al.}(2008)\citenamefont{Kuroki, Onari, Arita,
  Usui, Tanaka, Kontani, and Aoki}}]{KurokiPRL}
\bibnamefont{Kuroki,}\bibinfo{author}{\bibfnamefont{~K.,}
  }\bibnamefont{Onari,}\bibinfo{author}{\bibfnamefont{~S.,}
  }\bibnamefont{Arita,}\bibinfo{author}{\bibfnamefont{~R.,}
  }\bibnamefont{Usui,}\bibinfo{author}{\bibfnamefont{~H.,}
  }\bibnamefont{Tanaka,}\bibinfo{author}{\bibfnamefont{~Y.,}
  }\bibnamefont{Kontani,}\bibinfo{author}{\bibfnamefont{~H.,}
  }\bibnamefont{\&~}\bibnamefont{Aoki,}\bibinfo{author}{\bibfnamefont{~H.}
  } Unconventional pairing originating from the disconnected Fermi surfaces of
  superconducting LaFeAsO$_{1-x}$F$_{x}$. \bibinfo{journal}{{\it Phys. Rev. Lett.}}
  \textbf{\bibinfo{volume}{101}}, \bibinfo{pages}{087004}
  (\bibinfo{year}{2008}).

\bibitem[{\citenamefont{Mazin et~al.}(2008)\citenamefont{Mazin, Singh,
  Johannes, and Du}}]{MazinPRL}
\bibnamefont{Mazin,}\bibinfo{author}{\bibfnamefont{~I.~I.,}
  }\bibnamefont{Singh,}\bibinfo{author}{\bibfnamefont{~D.~J.,}
  }\bibnamefont{Johannes,}\bibinfo{author}{\bibfnamefont{~M.~D.,}
  }\bibnamefont{\&~}\bibnamefont{Du,}\bibinfo{author}{\bibfnamefont{~M.~H.}
  } Unconventional superconductivity with a sign reversal in the order
  parameter of LaFeAsO$_{1-x}$F$_{x}$. \bibinfo{journal}{{\it Phys. Rev. Lett.}}
  \textbf{\bibinfo{volume}{101}}, \bibinfo{pages}{057003}
  (\bibinfo{year}{2008}).

\bibitem[{\citenamefont{Chubukov et~al.}(2008)\citenamefont{Chubukov, Efremov,
  and Eremin}}]{ChubukovPRB2008}
\bibnamefont{Chubukov,}\bibinfo{author}{\bibfnamefont{~A.~V.,}
  }\bibnamefont{Efremov,}\bibinfo{author}{\bibfnamefont{~D.~V.,}
  }\bibnamefont{\&~}\bibnamefont{Eremin,}\bibinfo{author}{\bibfnamefont{~I.}
  } Magnetism, superconductivity, and pairing symmetry in iron-based
  superconductors. \bibinfo{journal}{{\it Phys. Rev. B}}
  \textbf{\bibinfo{volume}{78}}, \bibinfo{pages}{134512}
  (\bibinfo{year}{2008}).

\bibitem[{\citenamefont{Graser et~al.}(2009)\citenamefont{Graser, Maier,
  Hirschfeld, and Scalapino}}]{Scalapino2009NJP}
\bibnamefont{Graser,}\bibinfo{author}{\bibfnamefont{~S.,}
  }\bibnamefont{Maier,}\bibinfo{author}{\bibfnamefont{~T.,}
  }\bibnamefont{Hirschfeld,}\bibinfo{author}{\bibfnamefont{~P.,}
  }\bibnamefont{\&~}\bibnamefont{Scalapino,}\bibinfo{author}{\bibfnamefont{~D.}} 
  Near-degeneracy of several pairing channels in multiorbital models for the Fe pnictides. 
  \bibinfo{journal}{{\it New J. Phys.}} 
  \textbf{\bibinfo{volume}{11}}, \bibinfo{pages}{025016}
  (\bibinfo{year}{2009}).

\bibitem[{\citenamefont{Mukuda et~al.}(2014)\citenamefont{Mukuda, Engetsu,
  Yamamoto, Lai, Yashima, Kitaoka, Takemori, Miyasaka, and
  Tajima}}]{Mukuda2014PRB89_064511}
\bibnamefont{Mukuda,}\bibinfo{author}{\bibfnamefont{~H.,}
  }\bibnamefont{Engetsu,}\bibinfo{author}{\bibfnamefont{~F.,}
  }\bibnamefont{Yamamoto,}\bibinfo{author}{\bibfnamefont{~K.,}
  }\bibnamefont{Lai,}\bibinfo{author}{\bibfnamefont{~K.~T.,}
  }\bibnamefont{Yashima,}\bibinfo{author}{\bibfnamefont{~M.,}
  }\bibnamefont{Kitaoka,}\bibinfo{author}{\bibfnamefont{~Y.,}
  }\bibnamefont{Takemori,}\bibinfo{author}{\bibfnamefont{~A.,}
  }\bibnamefont{Miyasaka,}\bibinfo{author}{\bibfnamefont{~S.,}
  }\bibnamefont{\&~}\bibnamefont{Tajima,}\bibinfo{author}{\bibfnamefont{~S.}
  } Enhancement of superconducting transition temperature due to
  antiferromagnetic spin fluctuations in iron pnictides LaFe(As$_{1-x}$P$_{x}$)(O$_{1-y}$F$_{y}$): $^{31}$P-NMR
  studies. \bibinfo{journal}{{\it Phys. Rev. B}} \textbf{\bibinfo{volume}{89}},
  \bibinfo{pages}{064511} (\bibinfo{year}{2014}).

\bibitem[{\citenamefont{Platt et~al.}(2011)\citenamefont{Platt, Thomale, and
  Hanke}}]{PlattPRB2011}
\bibnamefont{Platt,}\bibinfo{author}{\bibfnamefont{~C.,}
  }\bibnamefont{Thomale,}\bibinfo{author}{\bibfnamefont{~R.,}
  }\bibnamefont{\&~}\bibnamefont{Hanke,}\bibinfo{author}{\bibfnamefont{~W.}
  } Superconducting state of the iron pnictide LiFeAs: A combined
  density-functional and functional-renormalization-group study.
  \bibinfo{journal}{{\it Phys. Rev. B}} \textbf{\bibinfo{volume}{84}},
  \bibinfo{pages}{235121} (\bibinfo{year}{2011}).


\bibitem[{\citenamefont{Kontani and Onari}(2010)}]{KontaniOnariPRL}
\bibnamefont{Kontani,}\bibinfo{author}{\bibfnamefont{~H.,}
  }\bibnamefont{\&~}\bibnamefont{Onari,}\bibinfo{author}{\bibfnamefont{~S.}
  } Orbital-fluctuation-mediated superconductivity in iron pnictides: Analysis
  of the five-orbital Hubbard-Holstein model. \bibinfo{journal}{{\it Phys. Rev. Lett.}} 
  \textbf{\bibinfo{volume}{104}}, \bibinfo{pages}{157001}
  (\bibinfo{year}{2010}).


\bibitem[{\citenamefont{Fernandes et~al.}(2014)\citenamefont{Fernandes,
  Chubukov, and Schmalian}}]{Fernandes2014}
\bibnamefont{Fernandes,}\bibinfo{author}{\bibfnamefont{~R.,}
  }\bibnamefont{Chubukov,}\bibinfo{author}{\bibfnamefont{~A.,}
  }\bibnamefont{\&~}\bibnamefont{Schmalian,}\bibinfo{author}{\bibfnamefont{~J.}}
   What drives nematic order in
  iron-based superconductors ?. \bibinfo{journal}{{\it Nature Phys.}}
  \textbf{\bibinfo{volume}{10}}, \bibinfo{pages}{97--104} (\bibinfo{year}{2014}).

\bibitem[{\citenamefont{Yin et~al.}(2011)\citenamefont{Yin, Haule, and
  Kotliar}}]{ZPYin}
\bibnamefont{Yin,}\bibinfo{author}{\bibfnamefont{~Z.~P.,}
  }\bibnamefont{Haule,}\bibinfo{author}{\bibfnamefont{~K.,}
  }\bibnamefont{\&~}\bibnamefont{Kotliar,}\bibinfo{author}{\bibfnamefont{~G.}} 
  Kinetic frustration and the nature of
  the magnetic and paramagnetic states in iron pnictides and iron
  chalcogenides. \bibinfo{journal}{{\it Nature Mater.}} \textbf{\bibinfo{volume}{10}},
  \bibinfo{pages}{932--935} (\bibinfo{year}{2011}).

\bibitem[{\citenamefont{Misawa et~al.}(2012)\citenamefont{Misawa, Nakamura, and
  Imada}}]{misawa2012}
\bibnamefont{Misawa,}\bibinfo{author}{\bibfnamefont{~T.,}
  }\bibnamefont{Nakamura,}\bibinfo{author}{\bibfnamefont{~K.,}
  }\bibnamefont{\&~}\bibnamefont{Imada,}\bibinfo{author}{\bibfnamefont{~M.}
  } $Ab$ $initio$ evidence for strong correlation associated with Mott proximity
  in iron-based superconductors. \bibinfo{journal}{{\it Phys. Rev. Lett.}}
  \textbf{\bibinfo{volume}{108}}, \bibinfo{pages}{177007}
  (\bibinfo{year}{2012}).

\bibitem[{\citenamefont{Ishida et~al.}(2009)\citenamefont{Ishida, Nakai, and
  Hosono}}]{IshidaJPSJReview}
\bibnamefont{Ishida,}\bibinfo{author}{\bibfnamefont{~K.,}
  }\bibnamefont{Nakai,}\bibinfo{author}{\bibfnamefont{~Y.,}
  }\bibnamefont{\&~}\bibnamefont{Hosono,}\bibinfo{author}{\bibfnamefont{~H.}
  } To what extent iron-pnictide new superconductors have been clarified: A
  progress report. \bibinfo{journal}{{\it J. Phys. Soc. Jpn.}}
  \textbf{\bibinfo{volume}{78}}, \bibinfo{pages}{062001}
  (\bibinfo{year}{2009}).

\bibitem[{\citenamefont{Stewart}(2011)}]{StewartRMP2011}
\bibnamefont{Stewart,}\bibinfo{author}{\bibfnamefont{~G.~R.}}
  Superconductivity in iron compounds. \bibinfo{journal}{{\it Rev. Mod. Phys.}}
  \textbf{\bibinfo{volume}{83}}, \bibinfo{pages}{1589--1652} (\bibinfo{year}{2011}).

\bibitem[{\citenamefont{Scalapino}(2012)}]{ScalapinoRMP}
\bibnamefont{Scalapino,}\bibinfo{author}{\bibfnamefont{~D.~J.}} 
  A common thread: The pairing interaction for unconventional superconductors.
  \bibinfo{journal}{{\it Rev. Mod. Phys.}} \textbf{\bibinfo{volume}{84}},
  \bibinfo{pages}{1383--1417} (\bibinfo{year}{2012}).

\bibitem[{\citenamefont{Platt et~al.}(2013)\citenamefont{Platt, Hanke, and
  Thomale}}]{PlattHankeThomale2013AdvPhys}
\bibnamefont{Platt,}\bibinfo{author}{\bibfnamefont{~C.,}
  }\bibnamefont{Hanke,}\bibinfo{author}{\bibfnamefont{~W.,}
  }\bibnamefont{\&~}\bibnamefont{Thomale,}\bibinfo{author}{\bibfnamefont{~R.}} 
  Functional renormalization group for multi-orbital Fermi surface instabilities. 
  \bibinfo{journal}{{\it Adv. Phys.}} 
  \textbf{\bibinfo{volume}{62}}, \bibinfo{pages}{453--562}
  (\bibinfo{year}{2013}).

\bibitem[{\citenamefont{Yang et~al.}(2009)\citenamefont{Yang, H\"uvonen, Nagel,
  R\~o\ om, Ni, Canfield, Bud'ko, Carbotte, and Timusk}}]{JYang2009PRL}
\bibnamefont{Yang,}\bibinfo{author}{\bibfnamefont{~J.,}
  }\bibnamefont{H\"uvonen,}\bibinfo{author}{\bibfnamefont{~D.,}
  }\bibnamefont{Nagel,}\bibinfo{author}{\bibfnamefont{~U.,}
  }\bibnamefont{R\~o\~om, }\bibinfo{author}{\bibfnamefont{~T.,}
  }\bibnamefont{Ni,}\bibinfo{author}{\bibfnamefont{~N.,}
  }\bibnamefont{Canfield,}\bibinfo{author}{\bibfnamefont{~P.~C.,}
  }\bibnamefont{Bud'ko,}\bibinfo{author}{\bibfnamefont{~S.~L.,}
  }\bibnamefont{Carbotte,}\bibinfo{author}{\bibfnamefont{~J.~P.,}
  }\bibnamefont{\&~}\bibnamefont{Timusk,}\bibinfo{author}{\bibfnamefont{~T.}
  } Optical spectroscopy of superconducting Ba$_{0.55}$K$_{0.45}$Fe$_{2}$As$_{2}$: Evidence for
  strong coupling to low-energy bosons. \bibinfo{journal}{{\it Phys. Rev. Lett.}}
  \textbf{\bibinfo{volume}{102}}, \bibinfo{pages}{187003}
  (\bibinfo{year}{2009}).


\bibitem[{\citenamefont{Qazilbash et~al.}(2009)\citenamefont{Qazilbash, Hamlin,
  Baumbach, Zhang, Singh, Maple, and Basov}}]{Qazilbash2009NP}
\bibnamefont{Qazilbash,}\bibinfo{author}{\bibfnamefont{~M.,}
  }\bibnamefont{Hamlin,}\bibinfo{author}{\bibfnamefont{~J.,}
  }\bibnamefont{Baumbach,}\bibinfo{author}{\bibfnamefont{~R.,}
  }\bibnamefont{Zhang,}\bibinfo{author}{\bibfnamefont{~L.,}
  }\bibnamefont{Singh,}\bibinfo{author}{\bibfnamefont{~D.~J.,}
  }\bibnamefont{Maple,}\bibinfo{author}{\bibfnamefont{~M.,}
  }\bibnamefont{\&~}\bibnamefont{Basov,}\bibinfo{author}{\bibfnamefont{~D.}
  } Electronic correlations in the iron pnictides. \bibinfo{journal}{{\it Nature Phys.}} 
  \textbf{\bibinfo{volume}{5}}, \bibinfo{pages}{647--650}
  (\bibinfo{year}{2009}).

\bibitem[{\citenamefont{Degiorgi}(2011)}]{Degiorgi2011NJP}
\bibnamefont{Degiorgi,}\bibinfo{author}{\bibfnamefont{~L.}} Electronic
  correlations in iron-pnictide superconductors and beyond: lessons learned
  from optics. \bibinfo{journal}{{\it New J. Phys.}}
  \textbf{\bibinfo{volume}{13}}, \bibinfo{pages}{023011}
  (\bibinfo{year}{2011}).

\bibitem[{\citenamefont{Terashima et~al.}(2010)\citenamefont{Terashima, Kimata,
  Kurita, Satsukawa, Harada, Hazama, Imai, Sato, Kihou, Lee
  et~al.}}]{Terashima2010JPSJ}
\bibnamefont{Terashima,}\bibinfo{author}{\bibfnamefont{~T.,}
  }\bibnamefont{Kimata,}\bibinfo{author}{\bibfnamefont{~M.,}
  }\bibnamefont{Kurita,}\bibinfo{author}{\bibfnamefont{~N.,}
  }\bibnamefont{Satsukawa,}\bibinfo{author}{\bibfnamefont{~H.,}
  }\bibnamefont{Harada,}\bibinfo{author}{\bibfnamefont{~A.,}
  }\bibnamefont{Hazama,}\bibinfo{author}{\bibfnamefont{~K.,}
  }\bibnamefont{Imai,}\bibinfo{author}{\bibfnamefont{~M.,}
  }\bibnamefont{Sato,}\bibinfo{author}{\bibfnamefont{~A.,}
  }\bibnamefont{Kihou,}\bibinfo{author}{\bibfnamefont{~K.,}
  }\bibnamefont{Lee,}\bibinfo{author}{\bibfnamefont{~C.-H.}
  }\bibnamefont{$et$~$al$.} Fermi surface and mass
  enhancement in KFe$_{2}$As$_{2}$ from de Haas--van Alphen effect measurements.
  \bibinfo{journal}{{\it J. Phys. Soc. Jpn.}}
  \textbf{\bibinfo{volume}{79}}, \bibinfo{pages}{053702}
  (\bibinfo{year}{2010}).

\bibitem[{\citenamefont{Nakai et~al.}(2008)\citenamefont{Nakai, Ishida,
  Kamihara, Hirano, and Hosono}}]{Nakai2008JPSJ}
\bibnamefont{Nakai,}\bibinfo{author}{\bibfnamefont{~Y.,}
  }\bibnamefont{Ishida,}\bibinfo{author}{\bibfnamefont{~K.,}
  }\bibnamefont{Kamihara,}\bibinfo{author}{\bibfnamefont{~Y.,}
  }\bibnamefont{Hirano,}\bibinfo{author}{\bibfnamefont{~M.,}
  }\bibnamefont{\&~}\bibnamefont{Hosono,}\bibinfo{author}{\bibfnamefont{~H.}
  } Evolution from itinerant antiferromagnet to unconventional superconductor
  with fluorine doping in LaFeAs(O$_{1-x}$F$_{x}$) revealed by $^{75}$As and $^{139}$La nuclear
  magnetic resonance. \bibinfo{journal}{{\it J. Phys. Soc. Jpn.}} 
  \textbf{\bibinfo{volume}{77}}, \bibinfo{pages}{3701}
  (\bibinfo{year}{2008}).

\bibitem[{\citenamefont{Anisimov et~al.}(2002)\citenamefont{Anisimov, Nekrasov,
  Kondakov, Rice, and Sigrist}}]{Anisimov_2002}
\bibnamefont{Anisimov,}\bibinfo{author}{\bibfnamefont{~V.,}
  }\bibnamefont{Nekrasov,}\bibinfo{author}{\bibfnamefont{~I.,}
  }\bibnamefont{Kondakov,}\bibinfo{author}{\bibfnamefont{~D.,}
  }\bibnamefont{Rice,}\bibinfo{author}{\bibfnamefont{~T.,}
  }\bibnamefont{\&~}\bibnamefont{Sigrist,}\bibinfo{author}{\bibfnamefont{~M.}} 
  Orbital-selective Mott-insulator
  transition in Ca$_{2-x}$Sr$_x$RuO$_{4}$. 
  \bibinfo{journal}{{\it Eur. Phys. J. B}} \textbf{\bibinfo{volume}{25}},
  \bibinfo{pages}{191--201} (\bibinfo{year}{2002}).

\bibitem[{\citenamefont{Koga et~al.}(2004)\citenamefont{Koga, Kawakami, Rice,
  and Sigrist}}]{KogaPRL2004}
\bibnamefont{Koga,}\bibinfo{author}{\bibfnamefont{~A.,}
  }\bibnamefont{Kawakami,}\bibinfo{author}{\bibfnamefont{~N.,}
  }\bibnamefont{Rice,}\bibinfo{author}{\bibfnamefont{~T.~M.,}
  }\bibnamefont{\&~}\bibnamefont{Sigrist,}\bibinfo{author}{\bibfnamefont{~M.}} 
  Orbital-selective Mott transitions in
  the degenerate Hubbard model. \bibinfo{journal}{{\it Phys. Rev. Lett.}}
  \textbf{\bibinfo{volume}{92}}, \bibinfo{pages}{216402}
  (\bibinfo{year}{2004}).

\bibitem[{\citenamefont{Pruschke and Bulla}(2005)}]{Pruschke_2005}
\bibnamefont{Pruschke,}\bibinfo{author}{\bibfnamefont{~T.,}
  }\bibnamefont{\&~}\bibnamefont{Bulla,}\bibinfo{author}{\bibfnamefont{~R.}
  } Hund's coupling and the metal-insulator transition in the two-band Hubbard
  model. \bibinfo{journal}{{\it Eur. Phys. J. B}}
  \textbf{\bibinfo{volume}{44}}, \bibinfo{pages}{217--224}
  (\bibinfo{year}{2005}).

\bibitem[{\citenamefont{Arita and Held}(2005)}]{Arita_2005}
\bibnamefont{Arita,}\bibinfo{author}{\bibfnamefont{~R.,} }
\bibnamefont{\&~}\bibnamefont{Held,}\bibinfo{author}{\bibfnamefont{~K.}} 
  Orbital-selective Mott-Hubbard transition in the two-band Hubbard model.
  \bibinfo{journal}{{\it Phys. Rev. B}} \textbf{\bibinfo{volume}{72}},
  \bibinfo{pages}{201102} (\bibinfo{year}{2005}).

\bibitem[{\citenamefont{de'Medici et~al.}(2005)\citenamefont{de'Medici,
  Georges, and Biermann}}]{Medici_2005}
\bibnamefont{de'Medici,}\bibinfo{author}{\bibfnamefont{~L.,}
  }\bibnamefont{Georges,}\bibinfo{author}{\bibfnamefont{~A.,}
  }\bibnamefont{\&~}\bibnamefont{Biermann,}
  \bibinfo{author}{\bibfnamefont{~S.} } Orbital-selective Mott transition in
  multiband systems: Slave-spin representation and dynamical mean-field theory.
  \bibinfo{journal}{{\it Phys. Rev. B}} \textbf{\bibinfo{volume}{72}},
  \bibinfo{pages}{205124} (\bibinfo{year}{2005}).

\bibitem[{\citenamefont{Ferrero et~al.}(2005)\citenamefont{Ferrero, Becca,
  Fabrizio, and Capone}}]{Ferrero_2005}
\bibnamefont{Ferrero,}\bibinfo{author}{\bibfnamefont{~M.,}
  }\bibnamefont{Becca,}\bibinfo{author}{\bibfnamefont{~F.,}
  }\bibnamefont{Fabrizio,}\bibinfo{author}{\bibfnamefont{~M.,}
  }\bibnamefont{\&~}\bibnamefont{Capone,}\bibinfo{author}{\bibfnamefont{~M.}
  } Dynamical behavior across the Mott transition of two bands with different
  bandwidths. \bibinfo{journal}{{\it Phys. Rev. B}} \textbf{\bibinfo{volume}{72}},
  \bibinfo{pages}{205126} (\bibinfo{year}{2005}).

\bibitem[{\citenamefont{Costi and Liebsch}(2007)}]{Costi_2007}
\bibnamefont{Costi,}\bibinfo{author}{\bibfnamefont{~T.~A.,}
  }\bibnamefont{\&~}\bibnamefont{Liebsch,}
  \bibinfo{author}{\bibfnamefont{~A.}} Quantum phase transition in the
  two-band Hubbard model. \bibinfo{journal}{{\it Phys. Rev. Lett.}}
  \textbf{\bibinfo{volume}{99}}, \bibinfo{pages}{236404}
  (\bibinfo{year}{2007}).

\bibitem[{\citenamefont{Jakobi et~al.}(2009)\citenamefont{Jakobi, Bl\"umer, and
  van Dongen}}]{Jakobi_2009}
\bibnamefont{Jakobi,}\bibinfo{author}{\bibfnamefont{~E.,}
  }\bibnamefont{Bl\"umer,}\bibinfo{author}{\bibfnamefont{~N.,}
  }\bibnamefont{\&~}\bibnamefont{Dongen, van}
  \bibinfo{author}{\bibfnamefont{~P.}} Orbital-selective Mott transitions in
  a doped two-band Hubbard model. \bibinfo{journal}{{\it Phys. Rev. B}}
  \textbf{\bibinfo{volume}{80}}, \bibinfo{pages}{115109}
  (\bibinfo{year}{2009}).


\bibitem[{\citenamefont{de' Medici et~al.}(2011)\citenamefont{de' Medici,
  Mravlje, and Georges}}]{Medici_2011}
\bibnamefont{de'Medici,}\bibinfo{author}{\bibfnamefont{~L.,}
  }\bibnamefont{Mravlje,}\bibinfo{author}{\bibfnamefont{~J.,}
  }\bibnamefont{\&~}\bibnamefont{Georges,}
  \bibinfo{author}{\bibfnamefont{~A.}} Janus-faced influence of Hund's rule
  coupling in strongly correlated materials. 
  \bibinfo{journal}{{\it Phys. Rev. Lett.}} \textbf{\bibinfo{volume}{107}}, \bibinfo{pages}{256401}
  (\bibinfo{year}{2011}).


\bibitem[{\citenamefont{Greger et~al.}(2013)\citenamefont{Greger, Kollar, and
  Vollhardt}}]{Greger_2013}
\bibnamefont{Greger,}\bibinfo{author}{\bibfnamefont{~M.,}
  }\bibnamefont{Kollar,}\bibinfo{author}{\bibfnamefont{~M.,}
  }\bibnamefont{\&~}\bibnamefont{Vollhardt,}\bibinfo{author}{\bibfnamefont{~D.}} 
  Emergence of a common energy scale close to the orbital-selective Mott transition. 
  \bibinfo{journal}{{\it Phys. Rev. Lett.}} \textbf{\bibinfo{volume}{110}}, \bibinfo{pages}{046403}
  (\bibinfo{year}{2013}).

\bibitem[{\citenamefont{Ishida and Liebsch}(2010)}]{IshidaLiebsch}
\bibnamefont{Ishida,}\bibinfo{author}{\bibfnamefont{~H.,}
  }\bibnamefont{\&~}\bibnamefont{Liebsch,}\bibinfo{author}{\bibfnamefont{~A.}} 
  Fermi-liquid, non-Fermi-liquid, and
  Mott phases in iron pnictides and cuprates. \bibinfo{journal}{{\it Phys. Rev. B}}
  \textbf{\bibinfo{volume}{81}}, \bibinfo{pages}{054513}
  (\bibinfo{year}{2010}).

\bibitem[{\citenamefont{Aichhorn et~al.}(2010)\citenamefont{Aichhorn, Biermann,
  Miyake, Georges, and Imada}}]{Aichhorn}
\bibnamefont{Aichhorn,}\bibinfo{author}{\bibfnamefont{~M.,}
  }\bibnamefont{Biermann,}\bibinfo{author}{\bibfnamefont{~S.,}
  }\bibnamefont{Miyake,}\bibinfo{author}{\bibfnamefont{~T.,}
  }\bibnamefont{Georges,}\bibinfo{author}{\bibfnamefont{~A.,}
  }\bibnamefont{\&~}\bibnamefont{Imada,}\bibinfo{author}{\bibfnamefont{~M.}
  } Theoretical evidence for strong correlations and incoherent metallic state
  in FeSe. \bibinfo{journal}{{\it Phys. Rev. B}} \textbf{\bibinfo{volume}{82}},
  \bibinfo{pages}{064504} (\bibinfo{year}{2010}).

\bibitem[{\citenamefont{Tamai et~al.}(2010)\citenamefont{Tamai, Ganin,
  Rozbicki, Bacsa, Meevasana, King, Caffio, Schaub, Margadonna, Prassides
  et~al.}}]{Tamai}
\bibnamefont{Tamai,}\bibinfo{author}{\bibfnamefont{~A.,}
  }\bibnamefont{Ganin,}\bibinfo{author}{\bibfnamefont{~A.~Y.,}
  }\bibnamefont{Rozbicki,}\bibinfo{author}{\bibfnamefont{~E.,}
  }\bibnamefont{Bacsa,}\bibinfo{author}{\bibfnamefont{~J.,}
  }\bibnamefont{Meevasana,}\bibinfo{author}{\bibfnamefont{~W.,}
  }\bibnamefont{King,}\bibinfo{author}{\bibfnamefont{~P. D.~C.,}
  }\bibnamefont{Caffio,}\bibinfo{author}{\bibfnamefont{~M.,}
  }\bibnamefont{Schaub,}\bibinfo{author}{\bibfnamefont{~R.,}
  }\bibnamefont{Margadonna,}\bibinfo{author}{\bibfnamefont{~S.,}
  }\bibnamefont{Prassides,}\bibinfo{author}{\bibfnamefont{~K.}
  }\bibnamefont{$et$~$al$.} Strong electron correlations
  in the normal state of the iron-based FeSe$_{0.42}$Te$_{0.58}$ superconductor
  observed by angle-resolved photoemission spectroscopy.
  \bibinfo{journal}{{\it Phys. Rev. Lett.}} \textbf{\bibinfo{volume}{104}},
  \bibinfo{pages}{097002} (\bibinfo{year}{2010}).


\bibitem[{\citenamefont{Yi et~al.}(2013)\citenamefont{Yi, Lu, Yu, Riggs, Chu,
  Lv, Liu, Lu, Cui, Hashimoto et~al.}}]{Yi_PRL}
\bibnamefont{Yi,}\bibinfo{author}{\bibfnamefont{~M.,}
  }\bibnamefont{Lu,}\bibinfo{author}{\bibfnamefont{~D.~H.,}
  }\bibnamefont{Yu,}\bibinfo{author}{\bibfnamefont{~R.,}
  }\bibnamefont{Riggs,}\bibinfo{author}{\bibfnamefont{~S.~C.,}
  }\bibnamefont{Chu,}\bibinfo{author}{\bibfnamefont{~J.-H.,}
  }\bibnamefont{Lv,}\bibinfo{author}{\bibfnamefont{~B.,}
  }\bibnamefont{Liu,}\bibinfo{author}{\bibfnamefont{~Z.~K.,}
  }\bibnamefont{Lu,}\bibinfo{author}{\bibfnamefont{~M.,}
  }\bibnamefont{Cui,}\bibinfo{author}{\bibfnamefont{~Y.-T.,}
  }\bibnamefont{Hashimoto,}\bibinfo{author}{\bibfnamefont{~M.}
  }\bibnamefont{$et$~$al$.} Observation of temperature-induced
  crossover to an orbital-selective Mott phase in A$_{x}$Fe$_{2-y}$Se$_{2}$ (A=K, Rb)
  superconductors. \bibinfo{journal}{{\it Phys. Rev. Lett.}}
  \textbf{\bibinfo{volume}{110}}, \bibinfo{pages}{067003}
  (\bibinfo{year}{2013}).

\bibitem[{\citenamefont{Lanat\`a et~al.}(2013)\citenamefont{Lanat\`a, Strand,
  Giovannetti, Hellsing, de' Medici, and Capone}}]{Lanata_PRB}
\bibnamefont{Lanat\`a,}\bibinfo{author}{\bibfnamefont{~N.,}
  }\bibnamefont{Strand,}\bibinfo{author}{\bibfnamefont{~H. U.~R.,}
  }\bibnamefont{Giovannetti,}\bibinfo{author}{\bibfnamefont{~G.,}
  }\bibnamefont{Hellsing,}\bibinfo{author}{\bibfnamefont{~B.,}
  }\bibnamefont{Medici,de' }\bibinfo{author}{\bibfnamefont{~L.,}
  }\bibnamefont{\&~}\bibnamefont{Capone,}\bibinfo{author}{\bibfnamefont{~M.}
  } Orbital selectivity in Hund's metals: The iron chalcogenides.
  \bibinfo{journal}{{\it Phys. Rev. B}} \textbf{\bibinfo{volume}{87}},
  \bibinfo{pages}{045122} (\bibinfo{year}{2013}).

\bibitem[{\citenamefont{Hardy et~al.}(2013)\citenamefont{Hardy, B\"ohmer, Aoki,
  Burger, Wolf, Schweiss, Heid, Adelmann, Yao, Kotliar et~al.}}]{Hardy_PRL}
\bibnamefont{Hardy,}\bibinfo{author}{\bibfnamefont{~F.,}
  }\bibnamefont{B\"ohmer,}\bibinfo{author}{\bibfnamefont{~A.~E.,}
  }\bibnamefont{Aoki,}\bibinfo{author}{\bibfnamefont{~D.,}
  }\bibnamefont{Burger,}\bibinfo{author}{\bibfnamefont{~P.,}
  }\bibnamefont{Wolf,}\bibinfo{author}{\bibfnamefont{~T.,}
  }\bibnamefont{Schweiss,}\bibinfo{author}{\bibfnamefont{~P.,}
  }\bibnamefont{Heid,}\bibinfo{author}{\bibfnamefont{~R.,}
  }\bibnamefont{Adelmann,}\bibinfo{author}{\bibfnamefont{~P.,}
  }\bibnamefont{Yao,}\bibinfo{author}{\bibfnamefont{~Y.~X.,}
  }\bibnamefont{Kotliar,}\bibinfo{author}{\bibfnamefont{~G.}
  }\bibnamefont{$et$~$al$.} Evidence of strong
  correlations and coherence-incoherence crossover in the iron pnictide
  superconductor ${\mathrm{KFe}}_{2}{\mathrm{As}}_{2}$. \bibinfo{journal}{{\it Phys.
  Rev. Lett.}} \textbf{\bibinfo{volume}{111}}, \bibinfo{pages}{027002}
  (\bibinfo{year}{2013}).

\bibitem[{\citenamefont{Li et~al.}(2014)\citenamefont{Li, Zhang, Liu, Ding, Wu,
  Wang, Wen, and Xiao}}]{Li_PRB}
\bibnamefont{Li,}\bibinfo{author}{\bibfnamefont{~W.,}
  }\bibnamefont{Zhang,}\bibinfo{author}{\bibfnamefont{~C.,}
  }\bibnamefont{Liu,}\bibinfo{author}{\bibfnamefont{~S.,}
  }\bibnamefont{Ding,}\bibinfo{author}{\bibfnamefont{~X.,}
  }\bibnamefont{Wu,}\bibinfo{author}{\bibfnamefont{~X.,}
  }\bibnamefont{Wang,}\bibinfo{author}{\bibfnamefont{~X.,}
  }\bibnamefont{Wen,}\bibinfo{author}{\bibfnamefont{~H.-H.,}
  }\bibnamefont{\&~}\bibnamefont{Xiao,}\bibinfo{author}{\bibfnamefont{~M.}
  } Mott behavior in K$_{x}$Fe$_{2-y}$Se$_{2}$ superconductors studied by pump-probe
  spectroscopy. \bibinfo{journal}{{\it Phys. Rev. B}} \textbf{\bibinfo{volume}{89}},
  \bibinfo{pages}{134515} (\bibinfo{year}{2014}).

\bibitem[{\citenamefont{Yu and Si}(2013)}]{Yu_2013}
\bibnamefont{Yu,}\bibinfo{author}{\bibfnamefont{~R.,} }
\bibnamefont{\&~}\bibnamefont{Si,}\bibinfo{author}{\bibfnamefont{~Q.}} 
  Orbital-selective Mott phase in multiorbital models for alkaline iron selenides K$_{1-x}$Fe$_{2-y}$Se$_{2}$,
  \bibinfo{journal}{{\it Phys. Rev. Lett.}} \textbf{\bibinfo{volume}{110}},
  \bibinfo{pages}{146402} (\bibinfo{year}{2013}).

\bibitem[{\citenamefont{de' Medici et~al.}(2014)\citenamefont{de' Medici,
  Giovannetti, and Capone}}]{MediciCapone}
\bibnamefont{de'Medici,}\bibinfo{author}{\bibfnamefont{~L.,}
  }\bibnamefont{Giovannetti,}\bibinfo{author}{\bibfnamefont{~G.,}
  }\bibnamefont{\&~}\bibnamefont{Capone,}\bibinfo{author}{\bibfnamefont{~M.}
  } Selective Mott physics as a key to iron superconductors.
  \bibinfo{journal}{{\it Phys. Rev. Lett.}} \textbf{\bibinfo{volume}{112}},
  \bibinfo{pages}{177001} (\bibinfo{year}{2014}).

\bibitem[{\citenamefont{Wang et~al.}(2014)\citenamefont{Wang, Schmidt, Fischer,
  Tsurkan, Greger, Vollhardt, Loidl, and Deisenhofer}}]{Wang_2014}
\bibnamefont{Wang,}\bibinfo{author}{\bibfnamefont{~Z.,}
  }\bibnamefont{Schmidt,}\bibinfo{author}{\bibfnamefont{~M.,}
  }\bibnamefont{Fischer,}\bibinfo{author}{\bibfnamefont{~J.,}
  }\bibnamefont{Tsurkan,}\bibinfo{author}{\bibfnamefont{~V.,}
  }\bibnamefont{Greger,}\bibinfo{author}{\bibfnamefont{~M.,}
  }\bibnamefont{Vollhardt,}\bibinfo{author}{\bibfnamefont{~D.,}
  }\bibnamefont{Loidl,}\bibinfo{author}{\bibfnamefont{~A.,}
  }\bibnamefont{\&~}\bibnamefont{Deisenhofer,}\bibinfo{author}{\bibfnamefont{~J.}} 
  Orbital-selective metal--insulator transition and gap formation above $T_{\rm c}$ 
  in superconducting Rb$_{1-x}$Fe$_{2-y}$Se$_{2}$.
  \bibinfo{journal}{{\it Nature Commun.}} \textbf{\bibinfo{volume}{5}}
  \bibinfo{pages}{3202} (\bibinfo{year}{2014}).



\bibitem[{\citenamefont{Kou et~al.}(2009)\citenamefont{Kou, Li, and
  Weng}}]{Kou_EPL}
\bibnamefont{Kou,}\bibinfo{author}{\bibfnamefont{~S.-P.,}
  }\bibnamefont{Li,}\bibinfo{author}{\bibfnamefont{~T.,} }
  \bibnamefont{\&~}\bibnamefont{Weng,}\bibinfo{author}{\bibfnamefont{~Z.-Y.}} 
  Coexistence of itinerant electrons and local moments in iron-based superconductors.
  \bibinfo{journal}{{\it Euro. Phys. Lett.}} \textbf{\bibinfo{volume}{88}},
  \bibinfo{pages}{17010} (\bibinfo{year}{2009}).

\bibitem[{\citenamefont{Hackl and Vojta}(2009)}]{Hackl_NJ}
\bibnamefont{Hackl,}\bibinfo{author}{\bibfnamefont{~A.,} }
  \bibnamefont{\&~}\bibnamefont{Vojta,}\bibinfo{author}{\bibfnamefont{~M.}} 
  Pressure-induced magnetic transition and volume collapse in FeAs superconductors: an
  orbital-selective Mott scenario. \bibinfo{journal}{{\it New J. Phys.}}
  \textbf{\bibinfo{volume}{11}}, \bibinfo{pages}{055064}
  (\bibinfo{year}{2009}).

\bibitem[{\citenamefont{Yin et~al.}(2010)\citenamefont{Yin, Lee, and
  Ku}}]{Yin_PRL}
\bibnamefont{Yin,}\bibinfo{author}{\bibfnamefont{~W.-G.,}
  }\bibnamefont{Lee,}\bibinfo{author}{\bibfnamefont{~C.-C.,} }
  \bibnamefont{\&~}\bibnamefont{Ku,}\bibinfo{author}{\bibfnamefont{~W.} } Unified picture for
  magnetic correlations in iron-based superconductors. \bibinfo{journal}{{\it Phys.
  Rev. Lett.}} \textbf{\bibinfo{volume}{105}}, \bibinfo{pages}{107004}
  (\bibinfo{year}{2010}).

\bibitem[{\citenamefont{Zhang et~al.}(2012)\citenamefont{Zhang, Lee, Lin, Wu,
  Jeschke, and Valent\'\i}}]{Zhang_PRB}
\bibnamefont{Zhang,}\bibinfo{author}{\bibfnamefont{~Y.-Z.,}
  }\bibnamefont{Lee,}\bibinfo{author}{\bibfnamefont{~H.,}
  }\bibnamefont{Lin,}\bibinfo{author}{\bibfnamefont{~H.-Q.,}
  }\bibnamefont{Wu,}\bibinfo{author}{\bibfnamefont{~C.-Q.,}
  }\bibnamefont{Jeschke,}\bibinfo{author}{\bibfnamefont{~H.~O.,}
  }\bibnamefont{\&~}\bibnamefont{Valent\'\i,}
  \bibinfo{author}{\bibfnamefont{~R.} } General mechanism for orbital
  selective phase transitions. \bibinfo{journal}{{\it Phys. Rev. B}}
  \textbf{\bibinfo{volume}{85}}, \bibinfo{pages}{035123}
  (\bibinfo{year}{2012}).


\bibitem[{\citenamefont{Miyake et~al.}(2010)\citenamefont{Miyake, Nakamura,
  Arita, and Imada}}]{miyake2010}
\bibnamefont{Miyake,}\bibinfo{author}{\bibfnamefont{~T.,}
  }\bibnamefont{Nakamura,}\bibinfo{author}{\bibfnamefont{~K.,}
  }\bibnamefont{Arita,}\bibinfo{author}{\bibfnamefont{~R.,}
  }\bibnamefont{\&~}\bibnamefont{Imada,}\bibinfo{author}{\bibfnamefont{~M.}
  } Comparison of ab initio low-energy models for LaFePO, LaFeAsO,
  BaFe$_2$As$_2$, LiFeAs, FeSe, and FeTe: Electron correlation and covalency.
  \bibinfo{journal}{{\it J. Phys. Soc. Jpn.}} \textbf{\bibinfo{volume}{79}},
  \bibinfo{pages}{044705} (\bibinfo{year}{2010}).


\bibitem[{\citenamefont{Nakamura et~al.}(2010)\citenamefont{Nakamura,
  Yoshimoto, Nohara, and Imada}}]{nakamura2010}
\bibnamefont{Nakamura,}\bibinfo{author}{\bibfnamefont{~K.,}
  }\bibnamefont{Yoshimoto,}\bibinfo{author}{\bibfnamefont{~Y.,}
  }\bibnamefont{Nohara,}\bibinfo{author}{\bibfnamefont{~Y.,}
  }\bibnamefont{\&~}\bibnamefont{Imada,}\bibinfo{author}{\bibfnamefont{~M.}
  } \textit{Ab initio} low-dimensional physics opened up by dimensional
  downfolding: Application to LaFeAsO. \bibinfo{journal}{{\it J. Phys. Soc. Jpn.}}
  \textbf{\bibinfo{volume}{79}}, \bibinfo{pages}{123708}
  (\bibinfo{year}{2010}).

\bibitem[{\citenamefont{Tahara and Imada}(2008)}]{TaharaVMC_Full}
\bibnamefont{Tahara,}\bibinfo{author}{\bibfnamefont{~D.,} }
 \bibnamefont{\&~}\bibnamefont{Imada,}\bibinfo{author}{\bibfnamefont{~M.}}
  Variational Monte Carlo method combined with quantum-number-projection and multi-variable optimization.
  \bibinfo{journal}{{\it J. Phys. Soc. Jpn.}} \textbf{\bibinfo{volume}{77}},
  \bibinfo{pages}{114701} (\bibinfo{year}{2008}).

\bibitem[{\citenamefont{Chu et~al.}(2010)\citenamefont{Chu, Analytis, De~Greve,
  McMahon, Islam, Yamamoto, and Fisher}}]{Chu2010}
\bibnamefont{Chu,}\bibinfo{author}{\bibfnamefont{~J.-H.,}
  }\bibnamefont{Analytis,}\bibinfo{author}{\bibfnamefont{~J.~G.,}
  }\bibnamefont{De~Greve,}\bibinfo{author}{\bibfnamefont{~K.,}
  }\bibnamefont{McMahon,}\bibinfo{author}{\bibfnamefont{~P.~L.,}
  }\bibnamefont{Islam,}\bibinfo{author}{\bibfnamefont{~Z.,}
  }\bibnamefont{Yamamoto,}\bibinfo{author}{\bibfnamefont{~Y.,}
  }\bibnamefont{\&~}\bibnamefont{Fisher,}\bibinfo{author}{\bibfnamefont{~I.~R.}} 
  In-plane resistivity anisotropy in
  an underdoped iron arsenide superconductor. \bibinfo{journal}{{\it Science}}
  \textbf{\bibinfo{volume}{329}}, \bibinfo{pages}{824--826} (\bibinfo{year}{2010}).

\bibitem[{\citenamefont{Kasahara et~al.}(2012)\citenamefont{Kasahara, Shi,
  Hashimoto, Tonegawa, Mizukami, Shibauchi, Sugimoto, Fukuda, Terashima,
  Nevidomskyy et~al.}}]{Kasahara2012}
\bibnamefont{Kasahara,}\bibinfo{author}{\bibfnamefont{~S.,}
  }\bibnamefont{Shi,}\bibinfo{author}{\bibfnamefont{~H.,}
  }\bibnamefont{Hashimoto,}\bibinfo{author}{\bibfnamefont{~K.,}
  }\bibnamefont{Tonegawa,}\bibinfo{author}{\bibfnamefont{~S.,}
  }\bibnamefont{Mizukami,}\bibinfo{author}{\bibfnamefont{~Y.,}
  }\bibnamefont{Shibauchi,}\bibinfo{author}{\bibfnamefont{~T.,}
  }\bibnamefont{Sugimoto,}\bibinfo{author}{\bibfnamefont{~K.,}
  }\bibnamefont{Fukuda,}\bibinfo{author}{\bibfnamefont{~T.,}
  }\bibnamefont{Terashima,}\bibinfo{author}{\bibfnamefont{~T.,}
  }\bibnamefont{Nevidomskyy,}\bibinfo{author}{\bibfnamefont{~A.~H.}
  }\bibnamefont{$et$~$al$.} Electronic nematicity above the
  structural and superconducting transition in BaFe$_{2}$(As$_{1-x}$P$_{x}$)$_{2}$. 
  \bibinfo{journal}{{\it Nature }}
  \textbf{\bibinfo{volume}{486}}, \bibinfo{pages}{382--385} (\bibinfo{year}{2012}).

\bibitem[{\citenamefont{Yu et~al.}(2014)\citenamefont{Yu, Zhu, and
  Si}}]{Yu_2014}
\bibnamefont{Yu,}\bibinfo{author}{\bibfnamefont{~R.,}
  }\bibnamefont{Zhu,}\bibinfo{author}{\bibfnamefont{~J.-X.,} }
  \bibnamefont{\&~}\bibnamefont{Si,}\bibinfo{author}{\bibfnamefont{~Q.}} Orbital-selective
  superconductivity, gap anisotropy, and spin resonance excitations in a
  multiorbital $t$-${J}_{1}$-${J}_{2}$ model for iron pnictides.
  \bibinfo{journal}{{\it Phys. Rev. B}} \textbf{\bibinfo{volume}{89}},
  \bibinfo{pages}{024509} (\bibinfo{year}{2014}).


\bibitem[{\citenamefont{Lang et~al.}(2010)\citenamefont{Lang, Grafe, Paar,
  Hammerath, Manthey, Behr, Werner, and B\"uchner}}]{LangPRL2010}
\bibnamefont{Lang,}\bibinfo{author}{\bibfnamefont{~G.,}
  }\bibnamefont{Grafe,}\bibinfo{author}{\bibfnamefont{~H.-J.,}
  }\bibnamefont{Paar,}\bibinfo{author}{\bibfnamefont{~D.,}
  }\bibnamefont{Hammerath,}\bibinfo{author}{\bibfnamefont{~F.,}
  }\bibnamefont{Manthey,}\bibinfo{author}{\bibfnamefont{~K.,}
  }\bibnamefont{Behr,}\bibinfo{author}{\bibfnamefont{~G.,}
  }\bibnamefont{Werner,}\bibinfo{author}{\bibfnamefont{~J.,}
  }\bibnamefont{\&~}\bibnamefont{B\"uchner,}\bibinfo{author}{\bibfnamefont{~B.}} Nanoscale electronic order in iron
  pnictides. \bibinfo{journal}{{\it Phys. Rev. Lett.}}
  \textbf{\bibinfo{volume}{104}}, \bibinfo{pages}{097001}
  (\bibinfo{year}{2010}).

\bibitem[{\citenamefont{Park et~al.}(2009)\citenamefont{Park, Inosov,
  Niedermayer, Sun, Haug, Christensen, Dinnebier, Boris, Drew, Schulz
  et~al.}}]{ParkPRL2009}
\bibnamefont{Park,}\bibinfo{author}{\bibfnamefont{~J.~T.,}
  }\bibnamefont{Inosov,}\bibinfo{author}{\bibfnamefont{~D.~S.,}
  }\bibnamefont{Niedermayer,}\bibinfo{author}{\bibfnamefont{~C.,}
  }\bibnamefont{Sun,}\bibinfo{author}{\bibfnamefont{~G.~L.,}
  }\bibnamefont{Haug,}\bibinfo{author}{\bibfnamefont{~D.,}
  }\bibnamefont{Christensen,}\bibinfo{author}{\bibfnamefont{~N.~B.,}
  }\bibnamefont{Dinnebier,}\bibinfo{author}{\bibfnamefont{~R.,}
  }\bibnamefont{Boris,}\bibinfo{author}{\bibfnamefont{~A.~V.,}
  }\bibnamefont{Drew,}\bibinfo{author}{\bibfnamefont{~A.~J.,}
  }\bibnamefont{Schulz,}\bibinfo{author}{\bibfnamefont{~L.}
  }\bibnamefont{$et$~$al$.} Electronic phase separation in
  the slightly underdoped iron pnictide superconductor Ba$_{1-x}$K$_{x}$Fe$_{2}$As$_{2}$.
  \bibinfo{journal}{{\it Phys. Rev. Lett.}} \textbf{\bibinfo{volume}{102}},
  \bibinfo{pages}{117006} (\bibinfo{year}{2009}).

\bibitem[{\citenamefont{Inosov et~al.}(2009)\citenamefont{Inosov, Leineweber,
  Yang, Park, Christensen, Dinnebier, Sun, Niedermayer, Haug, Stephens
  et~al.}}]{InosovPRB2009}
\bibnamefont{Inosov,}\bibinfo{author}{\bibfnamefont{~D.~S.,}
  }\bibnamefont{Leineweber,}\bibinfo{author}{\bibfnamefont{~A.,}
  }\bibnamefont{Yang,}\bibinfo{author}{\bibfnamefont{~X.,}
  }\bibnamefont{Park,}\bibinfo{author}{\bibfnamefont{~J.~T.,}
  }\bibnamefont{Christensen,}\bibinfo{author}{\bibfnamefont{~N.~B.,}
  }\bibnamefont{Dinnebier,}\bibinfo{author}{\bibfnamefont{~R.,}
  }\bibnamefont{Sun,}\bibinfo{author}{\bibfnamefont{~G.~L.,}
  }\bibnamefont{Niedermayer,}\bibinfo{author}{\bibfnamefont{~C.,}
  }\bibnamefont{Haug,}\bibinfo{author}{\bibfnamefont{~D.,}
  }\bibnamefont{Stephens,}\bibinfo{author}{\bibfnamefont{~P.~W.}
  }\bibnamefont{$et$~$al$.} Suppression of the structural
  phase transition and lattice softening in slightly underdoped Ba$_{1-x}$K$_{x}$Fe$_{2}$As$_{2}$  with electronic
  phase separation. \bibinfo{journal}{{\it Phys. Rev. B}}
  \textbf{\bibinfo{volume}{79}}, \bibinfo{pages}{224503}
  (\bibinfo{year}{2009}).

\bibitem[{\citenamefont{Li et~al.}(2012)\citenamefont{Li, Ding, Deng, Chang,
  Song, He, Wang, Ma, Hu, Chen et~al.}}]{Li_2012}
\bibnamefont{Li,}\bibinfo{author}{\bibfnamefont{~W.,}
  }\bibnamefont{Ding,}\bibinfo{author}{\bibfnamefont{~H.,}
  }\bibnamefont{Deng,}\bibinfo{author}{\bibfnamefont{~P.,}
  }\bibnamefont{Chang,}\bibinfo{author}{\bibfnamefont{~K.,}
  }\bibnamefont{Song,}\bibinfo{author}{\bibfnamefont{~C.,}
  }\bibnamefont{He,}\bibinfo{author}{\bibfnamefont{~K.,}
  }\bibnamefont{Wang,}\bibinfo{author}{\bibfnamefont{~L.,}
  }\bibnamefont{Ma,}\bibinfo{author}{\bibfnamefont{~X.,}
  }\bibnamefont{Hu,}\bibinfo{author}{\bibfnamefont{~J.-P.,}
  }\bibnamefont{Chen,}\bibinfo{author}{\bibfnamefont{~X.}
  }\bibnamefont{$et$~$al$.} Phase separation and magnetic
  order in K-doped iron selenide superconductor. 
  \bibinfo{journal}{{\it Nature Phys.}} \textbf{\bibinfo{volume}{8}}, \bibinfo{pages}{126--130}
  (\bibinfo{year}{2012}).

\bibitem[{\citenamefont{Texier et~al.}(2012)\citenamefont{Texier, Deisenhofer,
  Tsurkan, Loidl, Inosov, Friemel, and Bobroff}}]{Texier_2012}
\bibnamefont{Texier,}\bibinfo{author}{\bibfnamefont{~Y.,}
  }\bibnamefont{Deisenhofer,}\bibinfo{author}{\bibfnamefont{~J.,}
  }\bibnamefont{Tsurkan,}\bibinfo{author}{\bibfnamefont{~V.,}
  }\bibnamefont{Loidl,}\bibinfo{author}{\bibfnamefont{~A.,}
  }\bibnamefont{Inosov,}\bibinfo{author}{\bibfnamefont{~D.~S.,}
  }\bibnamefont{Friemel,}\bibinfo{author}{\bibfnamefont{~G.,}
  }\bibnamefont{\&~}\bibnamefont{Bobroff,} 
  \bibinfo{author}{\bibfnamefont{~J.,} } NMR study in the iron-selenide
  ${\mathrm{Rb}}_{0.74}{\mathrm{Fe}}_{1.6}{\mathrm{Se}}_{2}$: Determination of
  the superconducting phase as iron vacancy-free
  ${\mathrm{Rb}}_{0.3}{\mathrm{Fe}}_{2}{\mathrm{Se}}_{2}$,
  \bibinfo{journal}{Phys. Rev. Lett.} \textbf{\bibinfo{volume}{108}},
  \bibinfo{pages}{237002} (\bibinfo{year}{2012}).

\bibitem[{\citenamefont{Nomura et~al.}(2014)\citenamefont{Nomura, Nakamura, and
  Arita}}]{NomuraPRL}
\bibnamefont{Nomura,}\bibinfo{author}{\bibfnamefont{~Y.,}
  }\bibnamefont{Nakamura,}\bibinfo{author}{\bibfnamefont{~K.,}
  }\bibnamefont{\&~}\bibnamefont{Arita,}\bibinfo{author}{\bibfnamefont{~R.}
  } Effect of electron-phonon interactions on orbital fluctuations in
  iron-based superconductors. \bibinfo{journal}{{\it Phys. Rev. Lett.}}
  \textbf{\bibinfo{volume}{112}}, \bibinfo{pages}{027002}
  (\bibinfo{year}{2014}).

\bibitem[{\citenamefont{Kamihara et~al.}(2006)\citenamefont{Kamihara,
  Hiramatsu, Hirano, Kawamura, Yanagi, Kamiya, and Hosono}}]{Kamihara_LaFePO}
\bibnamefont{Kamihara,}\bibinfo{author}{\bibfnamefont{~Y.,}
  }\bibnamefont{Hiramatsu,}\bibinfo{author}{\bibfnamefont{~H.,}
  }\bibnamefont{Hirano,}\bibinfo{author}{\bibfnamefont{~M.,}
  }\bibnamefont{Kawamura,}\bibinfo{author}{\bibfnamefont{~R.,}
  }\bibnamefont{Yanagi,}\bibinfo{author}{\bibfnamefont{~H.,}
  }\bibnamefont{Kamiya,}\bibinfo{author}{\bibfnamefont{~T.,}
  }\bibnamefont{\&~}\bibnamefont{Hosono,}\bibinfo{author}{\bibfnamefont{~H.}
  }  Iron-based layered superconductor: LaOFeP. 
  \bibinfo{journal}{{\it J. Am. Chem. Soc}} \textbf{\bibinfo{volume}{128}},
  \bibinfo{pages}{10012-10013} (\bibinfo{year}{2006}).

\bibitem[{\citenamefont{Iimura et~al.}(2012)\citenamefont{Iimura, Matuishi,
  Sato, Hanna, Muraba, Kim, Kim, Takata, and Hosono}}]{Iimura2012}
\bibnamefont{Iimura,}\bibinfo{author}{\bibfnamefont{~S.,}
  }\bibnamefont{Matuishi,}\bibinfo{author}{\bibfnamefont{~S.,}
  }\bibnamefont{Sato,}\bibinfo{author}{\bibfnamefont{~H.,}
  }\bibnamefont{Hanna,}\bibinfo{author}{\bibfnamefont{~T.,}
  }\bibnamefont{Muraba,}\bibinfo{author}{\bibfnamefont{~Y.,}
  }\bibnamefont{Kim,}\bibinfo{author}{\bibfnamefont{~S.~W.,}
  }\bibnamefont{Kim,}\bibinfo{author}{\bibfnamefont{~J.~E.,}
  }\bibnamefont{Takata,}\bibinfo{author}{\bibfnamefont{~M.,}
  }\bibnamefont{\&~}\bibnamefont{Hosono,}\bibinfo{author}{\bibfnamefont{~H.}
  } Two-dome structure in electron-doped iron arsenide superconductors.
  \bibinfo{journal}{{\it Nature Commun.}} \textbf{\bibinfo{volume}{3}},
  \bibinfo{pages}{943} (\bibinfo{year}{2012}).

\bibitem[{\citenamefont{Fujiwara et~al.}(2013)\citenamefont{Fujiwara, Tsutsumi,
  Iimura, Matsuishi, Hosono, Yamakawa, and Kontani}}]{Fujiwara2013PRL}
\bibnamefont{Fujiwara,}\bibinfo{author}{\bibfnamefont{~N.,}
  }\bibnamefont{Tsutsumi,}\bibinfo{author}{\bibfnamefont{~S.,}
  }\bibnamefont{Iimura,}\bibinfo{author}{\bibfnamefont{~S.,}
  }\bibnamefont{Matsuishi,}\bibinfo{author}{\bibfnamefont{~S.,}
  }\bibnamefont{Hosono,}\bibinfo{author}{\bibfnamefont{~H.,}
  }\bibnamefont{Yamakawa,}\bibinfo{author}{\bibfnamefont{~Y.,}
  }\bibnamefont{\&~}\bibnamefont{Kontani,}\bibinfo{author}{\bibfnamefont{~H.}} 
  Detection of antiferromagnetic
  ordering in heavily doped LaFeAsO$_{1-x}$H$_{x}$ pnictide superconductors using
  nuclear-magnetic-resonance techniques. \bibinfo{journal}{{\it Phys. Rev. Lett.}}
  \textbf{\bibinfo{volume}{111}}, \bibinfo{pages}{097002}
  (\bibinfo{year}{2013}).

\bibitem[{\citenamefont{Yamaura et~al.}()\citenamefont{Yamaura, Iimura, Kumai,
  Hiraka, Ikeda, Ishikawa, Miao, Torii, Kamiyama, Otomo et~al.}}]{Yamaura}
\bibnamefont{Yamaura,}\bibinfo{author}{\bibfnamefont{~J.,}
  }\bibnamefont{Iimura,}\bibinfo{author}{\bibfnamefont{~S.,}
  }\bibnamefont{Kumai,}\bibinfo{author}{\bibfnamefont{~R.,}
  }\bibnamefont{Hiraka,}\bibinfo{author}{\bibfnamefont{~H.,}
  }\bibnamefont{Ikeda,}\bibinfo{author}{\bibfnamefont{~K.,}
  }\bibnamefont{Ishikawa,}\bibinfo{author}{\bibfnamefont{~Y.,}
  }\bibnamefont{Miao,}\bibinfo{author}{\bibfnamefont{~P.,}
  }\bibnamefont{Torii,}\bibinfo{author}{\bibfnamefont{~S.,}
  }\bibnamefont{Kamiyama,}\bibinfo{author}{\bibfnamefont{~T.,}
  }\bibnamefont{Otomo,}\bibinfo{author}{\bibfnamefont{~T.}
  }\bibnamefont{$et$~$al$.},
  \bibinfo{howpublished}{unpublished}.

\bibitem[{\citenamefont{Misawa and Imada}()}]{MisawaHubbard}
\bibnamefont{Misawa,}\bibinfo{author}{\bibfnamefont{~T.,} }\bibnamefont{\&~}
  \bibnamefont{Imada,}\bibinfo{author}{\bibfnamefont{~M.}}
  Origin of high-$T_c$ superconductivity in doped Hubbard models and their extensions 
  -- Roles of uniform charge fluctuations --.~
  \bibinfo{howpublished}{arXiv:1306.1434}.

\bibitem[{\citenamefont{Emery et~al.}(1990)\citenamefont{Emery, Kivelson, and
  Lin}}]{EmeryKivelson}
\bibnamefont{Emery,}\bibinfo{author}{\bibfnamefont{~V.~J.,}
  }\bibnamefont{Kivelson,}\bibinfo{author}{\bibfnamefont{~S.~A.}
  }\bibnamefont{\&~}\bibnamefont{Lin,}\bibinfo{author}{\bibfnamefont{~H.~Q.}
  } Phase separation in the $t$-$J$ model.
  \bibinfo{journal}{{\it Phys. Rev. Lett.}} \textbf{\bibinfo{volume}{64}},
  \bibinfo{pages}{475--478} (\bibinfo{year}{1990}).

\bibitem[{\citenamefont{Imada}(2005)}]{Imada_PRB}
\bibnamefont{Imada,}\bibinfo{author}{\bibfnamefont{~M.} } Universality
  classes of metal-insulator transitions in strongly correlated electron
  systems and mechanism of high-temperature superconductivity.
  \bibinfo{journal}{{\it Phys. Rev. B}} \textbf{\bibinfo{volume}{72}},
  \bibinfo{pages}{075113} (\bibinfo{year}{2005}).


\bibitem[{\citenamefont{Zhou et~al.}(2011)\citenamefont{Zhou, Kotliar, and
  Wang}}]{ZhouKotliar}
\bibnamefont{Zhou,}\bibinfo{author}{\bibfnamefont{~S.,}
  }\bibnamefont{Kotliar,}\bibinfo{author}{\bibfnamefont{~G.,}
  }\bibnamefont{\&~} \bibnamefont{Wang,} \bibinfo{author}{\bibfnamefont{~Z.}
  } Extended Hubbard model of superconductivity driven by charge fluctuations
  in iron pnictides. \bibinfo{journal}{{\it Phys. Rev. B}}
  \textbf{\bibinfo{volume}{84}}, \bibinfo{pages}{140505(R)}
  (\bibinfo{year}{2011}).


\bibitem[{\citenamefont{Wu et~al.}(2013)\citenamefont{Wu, Pelleg, Logvenov,
  Bollinger, Sun, Boebinger, Vanevi{\'c}, Radovi{\'c}, and
  Bo{\v{z}}ovi{\'c}}}]{WuBozovic}
\bibnamefont{Wu,}\bibinfo{author}{\bibfnamefont{~J.,}
  }\bibnamefont{Pelleg,}\bibinfo{author}{\bibfnamefont{~O.,}
  }\bibnamefont{Logvenov,}\bibinfo{author}{\bibfnamefont{~G.,}
  }\bibnamefont{Bollinger,}\bibinfo{author}{\bibfnamefont{~A.,}
  }\bibnamefont{Sun,}\bibinfo{author}{\bibfnamefont{~Y.,}
  }\bibnamefont{Boebinger,}\bibinfo{author}{\bibfnamefont{~G.,}
  }\bibnamefont{Vanevi{\'c},}\bibinfo{author}{\bibfnamefont{~M.,}
  }\bibnamefont{Radovi{\'c},}\bibinfo{author}{\bibfnamefont{~Z.,}
  }\bibnamefont{\&~}\bibnamefont{Bo{\v{z}}ovi{\'c},} 
  \bibinfo{author}{\bibfnamefont{~I.} } Anomalous independence of interface
  superconductivity from carrier density. \bibinfo{journal}{{\it Nature Mater.}}
  \textbf{\bibinfo{volume}{12}}, \bibinfo{pages}{877--881} (\bibinfo{year}{2013}).


\bibitem[{\citenamefont{Qing-Yan et~al.}(2012)\citenamefont{Qing-Yan, Zhi,
  Wen-Hao, Zuo-Cheng, Jin-Song, Wei, Hao, Yun-Bo, Peng, Kai
  et~al.}}]{Qing_2012}
\bibnamefont{Qing-Yan,}\bibinfo{author}{\bibfnamefont{~W.,}
  }\bibnamefont{Zhi,}\bibinfo{author}{\bibfnamefont{~L.,}
  }\bibnamefont{Wen-Hao,}\bibinfo{author}{\bibfnamefont{~Z.,}
  }\bibnamefont{Zuo-Cheng,}\bibinfo{author}{\bibfnamefont{~Z.,}
  }\bibnamefont{Jin-Song,}\bibinfo{author}{\bibfnamefont{~Z.,}
  }\bibnamefont{Wei,}\bibinfo{author}{\bibfnamefont{~L.,}
  }\bibnamefont{Hao,}\bibinfo{author}{\bibfnamefont{~D.,}
  }\bibnamefont{Yun-Bo,}\bibinfo{author}{\bibfnamefont{~O.,}
  }\bibnamefont{Peng,}\bibinfo{author}{\bibfnamefont{~D.,}
  }\bibnamefont{Kai,}\bibinfo{author}{\bibfnamefont{~C.~}
  }\bibnamefont{$et$~$al$.} Interface-induced
  high-temperature superconductivity in single unit-cell FeSe films on SrTiO$_{3}$.
  \bibinfo{journal}{{\it Chin. Phys. Lett.}} \textbf{\bibinfo{volume}{29}},
  \bibinfo{pages}{037402} (\bibinfo{year}{2012}).


\bibitem[{\citenamefont{Gutzwiller}(1963)}]{Gutzwiller}
\bibnamefont{Gutzwiller,}\bibinfo{author}{\bibfnamefont{~M.~C.} } Effect of
  correlation on the ferromagnetism of transition metals.
  \bibinfo{journal}{{\it Phys. Rev. Lett.}} \textbf{\bibinfo{volume}{10}},
  \bibinfo{pages}{159--162} (\bibinfo{year}{1963}).

\bibitem[{\citenamefont{Jastrow}(1955)}]{Jastrow}
\bibnamefont{Jastrow,}\bibinfo{author}{\bibfnamefont{~R.}} Many-body problem
  with strong forces. \bibinfo{journal}{{\it Phys. Rev.}}
  \textbf{\bibinfo{volume}{98}}, \bibinfo{pages}{1479--1484} (\bibinfo{year}{1955}).

\bibitem[{\citenamefont{Sorella}(2001)}]{Sorella_PRB2001}
\bibnamefont{Sorella,}\bibinfo{author}{\bibfnamefont{~S.}} Generalized
  Lanczos algorithm for variational quantum Monte Carlo.
  \bibinfo{journal}{{\it Phys. Rev. B}} \textbf{\bibinfo{volume}{64}},
  \bibinfo{pages}{024512} (\bibinfo{year}{2001}).

\bibitem[{\citenamefont{Hirsch}(1987)}]{Hirsch1987}
\bibnamefont{Hirsch,}\bibinfo{author}{\bibfnamefont{~J.~E.}}
  Antiferromagnetic singlet pairs, high-frequency phonons, and
  superconductivity. \bibinfo{journal}{{\it Phys. Rev. B}}
  \textbf{\bibinfo{volume}{35}}, \bibinfo{pages}{8726--8729} (\bibinfo{year}{1987}).

\end{thebibliography}
\end{document}